\documentclass[twocolumn]{aastex631}
\makeatletter
\let\frontmatter@title@above=\relax
\makeatother

\usepackage{amsmath,footmisc}
\usepackage{placeins}
\usepackage{gensymb}

\newcommand\CPA{\citep{Cardenas-Avendano:2022csp}}
\newcommand\CTA{\citet{Cardenas-Avendano:2022csp}}
\newcommand\CA{\cite{Cardenas-Avendano:2022csp}}
\usepackage{siunitx}

\begin{document}

\title{Multi-frequency models of black hole photon rings from low-luminosity accretion disks}

\author[0009-0008-3202-2972]{Tejahni Desire}
\correspondingauthor{Tejahni Desire}
\email{td6241@princeton.edu}
\affiliation{Department of Astrophysical Sciences, Princeton University, Princeton NJ 08544, USA}

\author[0000-0001-9528-1826]{Alejandro C\'ardenas-Avenda\~no}
\affiliation{Computational Physics and Methods (CCS-2) \& Center for Nonlinear Studies (CNLS), \\Los Alamos National Laboratory, Los Alamos NM 87545, USA}

\author[0000-0003-2966-6220]{Andrew Chael}
\affiliation{Princeton Gravity Initiative, Princeton University, Princeton NJ 08544, USA}

\begin{abstract}
    Images of black holes encode both astrophysical and gravitational properties. Detecting highly-lensed features in images can differentiate between these two effects. We present an accretion disk emission model coupled to the Adaptive Analytical Ray Tracing (\texttt{AART}) code that allows a fast parameter space exploration of black hole photon ring images produced from synchrotron emission from $10$ to $670$ GHz. As an application, we systematically study several disk models and compute their total flux density, average radii, and optical depth. The model parameters are chosen around fiducial values calibrated to general relativistic magnetohydrodynamic (GRMHD) simulations and observations of M87*. For the parameter space studied, we characterize the transition between optically thin and thick regimes and the frequency at which the first photon ring is observable. Our results highlight the need for careful definitions of photon ring radius in the image domain, as in certain models, the highly lensed photon ring is dimmer than the direct emission at certain angles. We find that at low frequencies, the ring radii are set by the electron temperature, while at higher frequencies, the magnetic field strength plays a more significant role, demonstrating how multi-frequency analysis can also be used to infer plasma parameters. Lastly, we show how our implementation can qualitatively reproduce multifrequency black hole images from GRMHD simulations when adding time-variability to our disk model through Gaussian random fields. This approach provides a new method for simulating observations from the Event Horizon Telescope (EHT) and the proposed Black Hole Explorer (BHEX) space mission.
\end{abstract}

\section{Introduction}

Imaging black holes has evolved from a purely theoretical pursuit~\citep{Bardeen_1974} to a highly active area of astrophysical research. Understanding the physics of event-horizon-scale emission around black holes remains challenging due to (i) a limited range of observational frequencies where imaging is possible, (ii) the extremely small angular sizes of nearby black holes and limited resolution from Earth-based Very Long Baseline Interferometry (VLBI), and (iii) uncertainties in physical models of the emitting region. Despite these challenges, the Event Horizon Telescope (EHT) collaboration, employing Earth-based VLBI at 1.3 mm wavelength (230 GHz) successfully provided the first images of the supermassive black holes  M87*~\citep{EHTM87I} and Sgr~A*~\citep{EHTSgrAI}, which have apparent angular sizes of a few times the projected Schwarzschild radius at millimeter wavelengths. 
    
The choice of millimeter-wavelength VLBI for EHT observations is motivated by the need to obtain the highest possible angular resolution \citep{EHTII,Raymond2024}. As important, however, are the properties of the synchrotron emitting plasma in the accretion flows surrounding Sgr~A* and M87*. In both sources, the spectral energy distributions (SEDs) peak around $1$~mm wavelength, where the flow transitions from optically thick to optically thin emission \citep{EHTMWL, EHTSgrAII}. Future space-based missions could improve the angular resolution of millimeter VLBI observations by adding a telescope in space that will observe at the same frequencies as the ground-based EHT array~\citep{Johnson:2024ttr}. Increasing black hole image resolution beyond Earth-based baselines could significantly enhance our understanding of these black holes and the emitting plasma that surrounds them, turning them into precision gravitational and astrophysical laboratories~\citep{Johnson:2024ttr,Lupsasca:2024xhq}.

The third main difficulty of horizon-scale observations--dealing with physical uncertainties in modeling the source--also presents its own complexities and opportunities for future development. Modeling emission close to supermassive black holes requires capturing the interplay of both astrophysical processes and strong gravitational phenomena. Accretion models must incorporate the physical conditions of the disk, where density, temperature, and magnetic field strength play critical roles in shaping the distribution and characteristics of the emitted radiation. The plasma around Sgr~A* and M87* is hot, dilute, and highly magnetized, with its millimeter SED dominated by relativistic synchrotron emission  \citep{Yuan_Narayan_2014, EHTM87VIII, EHTSgrAVIII}. The accretion flows around both objects have been studied extensively with both analytic models \citep[e.g.,][]{Quataert99,Falcke2000,BroderickLoeb2006,Ozel2021} and General Relativistic Magnetohydrodynamic (GRMHD) simulations \citep[e.g.,][]{Moscibrodzka_09,Dexter_2012,Ressler17,Chael_19}. 

\begin{figure}[ht!]
        \plotone{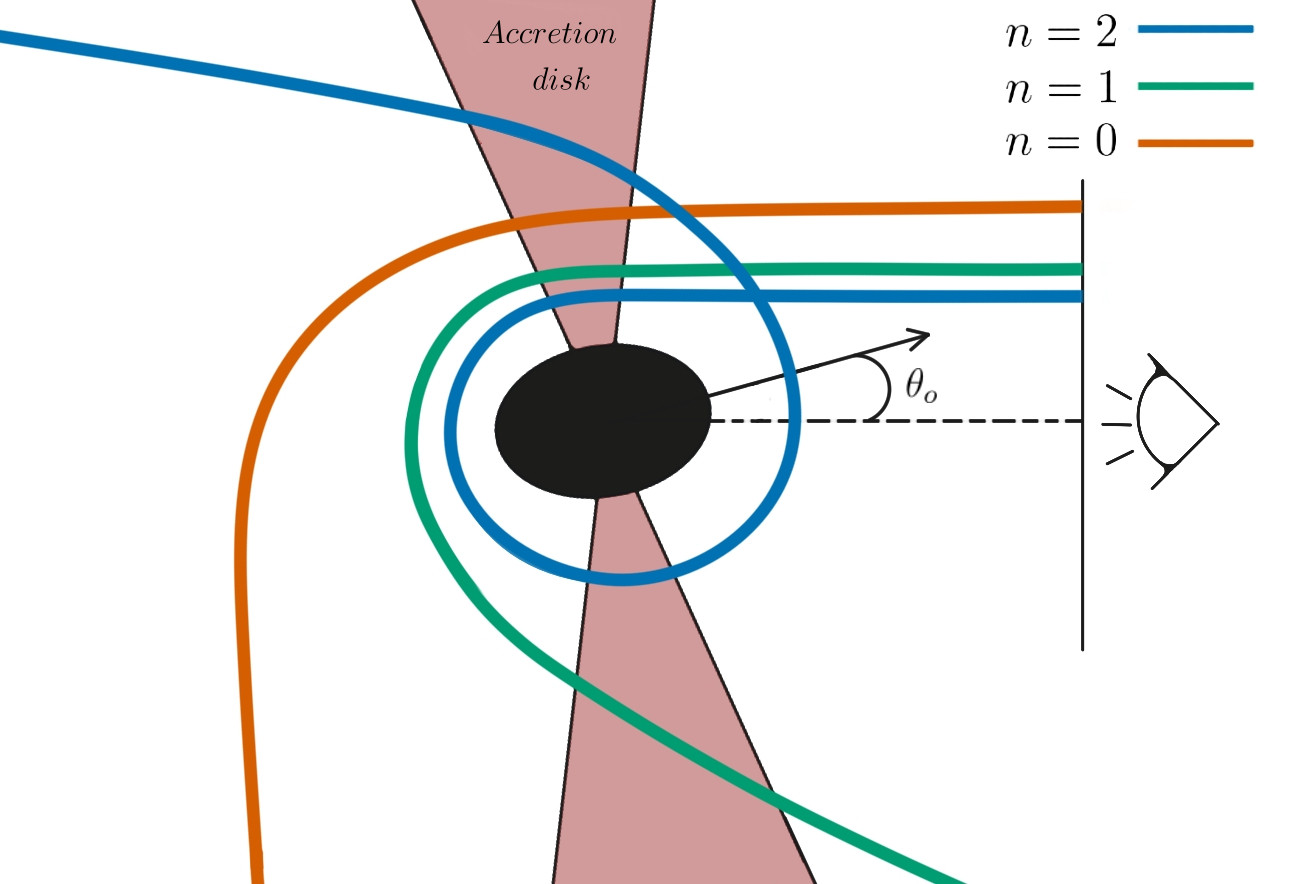}
        \vspace{12pt}
        \caption{Illustration of the lensing behavior that creates the observed photon rings around a black hole. In this model, emission is confined to an accretion disk (in red); the observation angle $\theta_{\rm o}$ is the angle between the normal of the disk and the observer's line of sight. The $n$th photon ring includes all photons that have passed through the black hole's equatorial plane $n+1$ times before reaching the observer.}
        \label{fig:photonRings}
    \end{figure}

Close to a black hole, gravitational redshift also affects the spectral features of the emitted light, and the spacetime geometry dictates the path of these photons to the observer~\citep{Bardeen_1974}. In an optically thin medium, unscattered photon trajectories, fully dictated by the theory of gravity, lead to the formation of photon rings--distinct features in the image plane created by extreme gravitational lensing \citep{JMD2020}. Photon rings can be labeled by counting the number of $n+1$ half-orbits the corresponding trajectories make around the black hole before reaching the observer (see, e.g.,~\citet{Cardenas-Avendano:2023obg,Wong:2024gph} for alternative definitions). In this definition, the $n=0$ direct image is the direct image formed by photons whose trajectories may still be highly lensed relative to a flat space model, but that made only a single pass through the equatorial plane in their journey to the observer. The first photon ring, $n=1$, is formed by the photons that make two passes through the equatorial plane (as illustrated in \autoref{fig:photonRings}). If the emitting plasma is optically thin, the final sky image will be the sum of the direct image ($n=0$), and the $n=1$ to $n=\infty$ photon rings~\citep{JMD2020}. As $n\rightarrow\infty$, the successive photon ring images become exponentially narrower, with their position converging to the image of the unstable photon shell, the so-called black hole shadow~\citep[or critical curve;][]{Bardeen_1974,JMD2020}.

According to General Relativity, the shape of photon rings with higher values of $n\geq2$ is (exponentially) dominated by the spacetime geometry, rather than by the astrophysical processes that generate the emission~\citep{JMD2020,GrallaLupsascaMarrone2020}. The shape and size of lower order photon rings depends both on the geometry and the astrophysics of the disk, with the $n=0$ image being most sensitive to the detailed model of the emitting material. Consequently, a precise measurement of the shape of the first photon ring can provide information about \emph{both} the black hole's astrophysical environment and the nature of the spacetime in its vicinity~\citep{JMD2020,Cardenas-Avendano:2023dzo,Lupsasca:2024xhq}. 
    
GRMHD simulations~\citep{EHTM87V,EHTSgrAI} are currently the best tool we have to study the competing effects of the astrophysical source and gravitational redshift/lensing on black hole images. Given a set of initial conditions and known conservation laws for plasma and magnetic fields, these simulations numerically evolve the system, which is then ray-traced to produce an image. The parameters of the simulation and radiative transfer are then tuned to match the EHT data. This data analysis procedure aims to infer the underlying parameters of the black hole, accretion disk, and jet by comparing simulated GRMHD images to data~\citep{EHTM87V}. With existing EHT data, measuring the geometric parameters of the source, such as the spin of the black hole and its inclination relative to the observer, remains challenging \citep{EHTM87VIII, EHTSgrAVIII}. 

State-of-the-art GRMHD simulations are highly computationally intensive, which limits the number of outputs and parameters that can be studied when analyzing observations~\citep{Wong:2022rqr}. To address this limitation, phenomenological approximations to the equations governing accretion disks have been developed as complementary models \citep[e.g.,][]{BroderickLoeb2006,2009ApJ...697.1164B,2021ApJ...912...35N}. One such phenomenological model to simulate black hole images is the Adaptive Analytical Ray Tracing (\texttt{AART}) code~\CPA. This code allows for fast simulation of black hole images that appear qualitatively similar to GRMHD simulations. The variability in \texttt{AART} is modeled using \texttt{inoisy}~\citep{Lee_Gammie_2021}, which generates realizations of Gaussian random fields with (prescribed and tuneable) position-dependent correlations. The ray tracing is performed entirely analytically on adaptive grids, which allocate more pixels in the image plane to accurately resolve the photon rings.

So far, \texttt{AART} has used phenomenological emission profiles rather than physical models of the accretion disk. These phenomenological models allow control over key features such as the peak, width, and decay of the emission, which is particularly useful for studying the relationship between spacetime geometry and the photon ring structure in a way that is independent of specific astrophysical assumptions~\citep{Cardenas-Avendano:2024flu}. However, while this generality is advantageous for probing the underlying geometry, it also limits the code’s ability to capture details of the astrophysical processes that govern emission mechanisms in real accretion disks.
    
Another significant limitation of a purely phenomenological approach is the inability to meaningfully vary the frequency of the produced images. This restriction limits the \texttt{AART} model's applicability for simulating the full range of observational data expected from upcoming observations~\citep{Chael_Issaoun_Pesce_Johnson_Ricarte_Fromm_Mizuno_2023,Johnson2023,Palumbo:2024jtz}, in particular the appearance of photon rings from the same emitting material at multiple frequencies. In addition, the current \texttt{AART}'s emission model assumes that the emitting region is completely optically thin. However, both Sgr~A* and M87* have SEDs that peak at wavelengths $\sim1$mm \citep{EHTMWL, EHTSgrAII}, indicating that the accretion flow transitions from optically thick to optically thin in the same frequency regime probed by the EHT. Even when the bulk of the flow is optically thin, the visibility of higher order photon rings can be ``cut off'' by a locally high optical depth as the photons' affine path length increases when orbiting the black hole multiple times~\citep{Palumbo:2024jtz}.

This work implements an astrophysically motivated emission model in \texttt{AART},\footnote{
The code used for these simulations is publicly available at~\url{https://github.com/iAART/AstroModels} and is preserved on Zenodo~\cite{10.5281/zenodo.14629536}.
} based on~\citet{2009ApJ...697.1164B} and~\citet{Dexter_2016}. While still simple, analytical, and computationally efficient, the model introduces three key variables: electron temperature, magnetic field strength, and particle density. In this work, we systematically study several disk models, covering a wide parameter space, chosen after calibrating to GRMHD simulations and observational constraints. For all the studied models, we compute their total flux, photon ring radii, and optical depth. We also characterize the transition between optically thin and thick regimes, and the frequency at which the first photon ring is observable.

The remainder of the paper is organized as follows. In \autoref{methods}, we define and implement an accretion model based on the full solution of the radiative transfer equation, including optical depth effects following~\citet{2009ApJ...697.1164B} and~\citet{Dexter_2016}. We then use this model to generate high-resolution, time-averaged black hole images and measure their photon ring radii (in the image plane), flux, and optical depth, along with their dependence on geometric and astrophysical parameters. In \autoref{sec:noisyblurr}, we leverage the full capabilities of \texttt{AART} to produce black hole movies by incorporating stochastic variability and a general accretion flow with arbitrary particle four-velocities. The resulting simulations are qualitatively similar to GRMHD snapshots, and once blurred, they resemble those captured by the EHT. Lastly, in~\autoref{sec:conclusion}, we discuss the implications and limitations of our results and outline directions for future research.
        
Throughout the equations, calligraphic variables, e.g., $\mathcal{I}$, represent relativistically invariant quantities (corresponding to those measured in the lab frame), while non-calligraphic variables, unless explicitly stated, denote those in the fluid frame. Unless otherwise specified, we adopt geometrical units where $G=c=1$, and the gravitational radius, $r_{\rm g}=GM/c^2$, where $M$ is the black hole mass, to measure distances.

% __________________________________________________________________________________________________________________________

\section{Astrophysical Model}
\label{methods}

Creating simulated images of black holes involves calculating the paths of photons in curved spacetime and then solving the radiative transfer equation for each photon's path. A full solution to the radiative transfer equation would account for emission, absorption, and scattering by different processes as the photons interact with matter in the accretion flow. In the millimeter wavelengths probed by the EHT, the radiation from low luminosity AGNs is dominated by synchrotron emission, and the radiative transfer equation can be solved along geodesics without accounting for scattering. 
For example, when producing images of synchrotron radiation from GRMHD simulations, codes like \texttt{ipole} \citep{Mościbrodzka_Gammie_2018} numerically solve the polarized radiative transfer equation for synchrotron emission, after interpolating properties of the simulation plasma and magnetic field at discretized points along geodesics computed from the observer backwards into the accretion flow. The result provides the intensity and synchrotron spectrum at each point in the image.

We adopt a simplified approach in our semi-analytic model for a synchrotron-emitting equatorial accretion disk around a supermassive black hole. Namely, we assume that the synchrotron emission and absorption coefficients are \emph{constant} on each pass of a geodesic through the disk, and thus we are able to analytically solve for the emergent radiation on each pass. This is an extension of the phenomenological approach proposed in~\citet{GrallaLupsascaMarrone2020}, but extended to include both synchrotron self-absorption and emission and absorption coefficients for thermal synchrotron radiation.
    
For each pixel $(\alpha,\beta)$, on the image screen, we use \texttt{AART} to compute the photon geodesics analytically backwards into the spacetime. In particular, at each point we compute the maximum number of sub-images, $n_{\max}(\alpha,\beta)$, contributing to the emission in each pixel; the total number of passes through the equatorial disk is given by $n_{\max}$+1. For the purpose of this work, we have solved the radiative transfer equation only including contributions through the second photon ring.\footnote{Millimeter images of M87* are thought to only be able to resolve up to the $n=2$ layer even with space-based millimeter VLBI \citep{JMD2020,GrallaLupsascaMarrone2020}.} The \texttt{AART} image consists of grids of variable resolution that allows it to increase the number of pixels in regions with larger $n_{\max}$, effectively concentrating resolution in the highly-lensed photon rings and making the calculation of high-resolution images very efficient~\CPA. 
    
Instead of computing physical synchrotron emission and absorption coefficients from an accretion disk model, \texttt{AART} originally used a phenomenological emission profile \citep[][Eq. 6]{GrallaLupsascaMarrone2020} in a strict optically thin assumption. This phenomenological approach allowed for a wide exploration of the effects of gravitational lensing on image structures, but it did not allow \texttt{AART} to be used to predict images from physical models for the accretion disk, or to explore how different profiles of particle density, magnetic field strength, and temperature might change the observed images. As a consequence, the phenomenological emission profile used in~\citet{GrallaLupsascaMarrone2020} and \CTA~says nothing about how the emitted or absorbed radiation changes with the observation frequency. The ability to model and interpret multi-frequency images of black hole accretion flows will be essential in the near future as the EHT gains access to observing at both $230$ and $345$ GHz \citep{Johnson2023,Chael_Issaoun_Pesce_Johnson_Ricarte_Fromm_Mizuno_2023}. 

\subsection{The Radiative Transfer Equation}
\label{sec:RTE}

The evolution of the photon's intensity along the geodesic from the disk to the observer is captured in the radiative transfer process, where energy is either lost through redshift or exchanged with the disk plasma via absorption and emission, which add or remove photons from the null ray. In addition to these standard radiative transfer effects, general relativistic radiative transfer must correctly account for both the bending of the photon' trajectories in curved spacetime, as well as the loss/gain of photon energy from general and special relativistic red/blueshift. The gravitational redshift $g$ is given by 
    \begin{align}
    \label{Eq:RedshiftDefinition}
         g=\frac{\nu_0}{\nu}=\frac{k_t}{k_\mu u^\mu},
     \end{align}
where $\nu$ denotes the frequency of the emitted radiation, $\nu_0$ is the observed frequency, $k^\mu$ is the photon wave vector, and $u^\mu$ is the four-velocity of the emitting plasma. 

The change of specific intensity $I_\nu$ with respect to the affine parameter $\lambda$ along the geodesic, depends on the local absorption coefficient $\kappa_\nu$, and the emission coefficient $j_\nu$, through the general relativistic radiative transfer equation (RTE; e.g.,~\citealt{Dexter_2016}):
    \begin{align}
        \frac{d\mathcal{I}_{\nu}}{d\lambda} &= -\mathcal{K}_{\nu} \mathcal{I}_{\nu} + \mathcal{J}_{\nu} \label{eq:Raddiffeq},        
    \end{align}
where the rest frame specific intensity and the absorption/emission coefficients are multiplied by factors of the redshift to create these relativistically invariant quantities:
    \begin{align}
        \mathcal{I}_{\nu} &= g^3I_\nu,\nonumber\\ \mathcal{K}_{\nu} &= g^{-1}\kappa_\nu,\nonumber\\
        \mathcal{J}_{\nu} &= g^2 j_\nu. \label{eq:invarientJ}
    \end{align}
Similarly to the non-relativistic version, the general relativistic radiative transfer equation can be rewritten in terms of the source function $S_\nu = j_\nu / \kappa_\nu$, and optical depth $\tau_\nu = \int \kappa_\nu(\lambda')d\lambda'$ as 
    \begin{align}
        \frac{d\mathcal{I}_{\nu}}{d\tau_{\nu}} &= -\mathcal{I}_{\nu} + \mathcal{S}_{\nu} \label{eq:diffeq},   
    \end{align}
where the source function transforms like the intensity, $\mathcal{S}_{\nu} = g^3 S_{\nu}$ \citep{Dexter_2016,Gammie_Leung_2012}. Notably, the optical depth $\tau_\nu$ is a relativistic invariant \citep{Rybicki:847173}.

Our model makes the simplifying assumption that $j_\nu$ and $\kappa_\nu$ are constant in each individual pass through the equatorial disk. As a result, we can solve the RTE, \autoref{eq:diffeq}, analytically. This assumption implies that the final intensity measured by an observer can be decomposed in terms of the contribution made by each of the $n+1$ passes of the photon through the disk. We define the $n$th sub-image intensity as 
    \begin{align}
        \mathcal{I}_{\nu_0,n} &= \mathcal{S}_{\nu,n} (1 - e^{-\tau_{\nu,n}}) \label{eq:simpleAbsorb},
    \end{align}
i.e., $\mathcal{I}_{\nu_0,n}$ is the intensity produced in a single pass through the disk. For all $n>0$, this intensity is diminished by the cumulative optical depth of the subsequent passes through the disk. In \autoref{eq:simpleAbsorb}, the frequency of the emitted radiation is written in terms of the local redshift $g$ at the location where the geodesic intersects the disk using \autoref{Eq:RedshiftDefinition}, i.e., $\nu=\nu_0/g$.
    The total intensity at the observed frequency $\nu_0$ on the observer image at the pixel location $(\alpha,\beta)$ is thus given by adding up all contributions as follows
    \begin{equation}
        \mathcal{I}_{\nu_0}(\alpha,\beta) = \sum_{n=0}^{n_{\max}} \mathcal{S}_{\nu,n}\left(1-e^{-\tau_{\nu,n}}\right)e^{-\left(\sum_{i=0}^{n-1}\tau_{\nu,i}\right)}\label{eq:simpleAbsorb2}.
    \end{equation}
In this expression, the $n$th photon ring intensity is exponentially diminished through the optical depth of all previous passes $\tau_{\nu,n}$. Note that we solve \autoref{eq:simpleAbsorb2} recursively \emph{backwards} from the most-lensed emission $n_{\max}$, where we assume $\mathcal{I}_n=0$ for all $n>n_{\max}$.
 
The optical depth for each pass through the disk is given by the local absorption coefficient $\kappa_\nu$ and path length $\ell$ through the disk in the fluid rest frame:  
    \begin{align}
        \tau_{\nu} &= \kappa_\nu \ell, \\
        \ell &= \frac{r  \theta_{\rm disk}}{\cos(\theta_e)}. \label{eq:opticalDepth}
    \end{align}
In \autoref{eq:opticalDepth}, the disk opening angle $\theta_{\rm disk}$ denotes the ratio of the local disk height $h$ to disk radius $r$. The incident angle $\theta_e$ is the angle between the geodesic and the normal to the disk in the fluid frame, defined by~\citep{Cunningham1975}
    \begin{equation}
        \cos(\theta_e) = g\frac{\sqrt{\eta}}{r} ,
    \end{equation}
    where the Carter constant $\eta$ is a conserved quantity related to the photons angular momentum, given by~\citep{Bardeen_1974}
    \begin{align}
        \eta &= (\alpha^2-a_*^2)\cos^2(\theta_{\rm o})+\beta^2,
    \end{align}
for an image pixel $(\alpha,\beta)$, viewer inclination angle $\theta_{\rm o}$, and black hole spin $a_*$. 

In the optically thin regime, the optical depth is relatively small compared to the source function $S_\nu$, and therefore one can neglect $\kappa_\nu$ in \autoref{eq:Raddiffeq}. Thus, in this regime, also assuming $\mathcal{J}_{\nu}$ is constant through each individual pass through the disk, the intensity contribution of the $n$th photon ring will be given simply by
\begin{align}
        \mathcal{I}_{\nu_0,n} &= \mathcal{J}_{\nu,n}\ell'_n,\label{eq:thinSol}
\end{align}
where $\ell'_n=g\ell_n$ is the lab frame path length of the $n$th pass through the disk. When neglecting the optical depth, each photon ring's intensity is independent from the others, allowing the final image to be computed by simply summing each layer as 
\begin{align}
\mathcal{I}_{\nu_0}(\alpha,\beta) &= \sum_{n=0}^{n_{\rm max}}\mathcal{J}_{\nu,n}\,\ell'_n.\label{eq:thinSolTotal}
\end{align}

In this model we also assume that there are no major effects from inverse Compton scattering in the interim between emission and observation. This assumption is well-justified for this work, as this particular contribution is only expected to become significant at optical and X-ray frequencies~\citep{Prather_Dexter}.
    
\subsection{Synchrotron radiation \label{sec:sync}}
    
We model the disk's emission as synchrotron radiation produced by thermal electrons. As we focus our frequency range around the current observation frequency of the EHT at $230$GHz, we do not include non-thermal electrons, as they are thought to only be relevant to measured intensities at lower frequencies in the accretion disk or extended jet \citep[e.g.,][]{2009ApJ...697.1164B,EHTM87I}. 
    
The thermal synchrotron emission coefficient $j_\nu$ is given by~\citep{Dexter_2016},
    \begin{align}
        j_\nu &=\frac{n_{th}e^{2}\nu}{2\sqrt{3}\,c\,\Theta_{e}^{2}}\;I_{I}\left(\frac{\nu}{\nu_c} \right), 
        \label{eq:jemitI}
    \end{align}
    where $\nu$ denotes the emitted frequency, $n_{th}$ is the electron number density, and $\Theta_e = (k_B T_e)/(m_e{c}^{2})$ is the dimensionless electron temperature. The electron temperature (in Kelvin) is denoted by $T_e$, $k_B$ is Boltzmann's constant, $e$ is the electron charge, and $m_e$ is the electron mass. The critical synchrotron frequency $\nu_c$ is defined as
    \begin{equation}
        \nu_c=\frac{3}{4\pi}\frac{eB\Theta_e^{2}}{m_ec}\sin\theta_B,
        \label{eq:nucrit}
    \end{equation}
    where $B$ is the local magnetic field strength and $\theta_B$ is the local angle between the magnetic field and the wave vector. This angle is defined in the emitter's rest frame as~\citep{Dexter_2016}
    \begin{align}
        \cos^2{\theta_B} &= \frac{\left(b^\mu k_\mu \right)^2}{b^\nu b_\nu \left(k^\sigma u_\sigma \right)}, \label{eq:cosang} 
    \end{align}
    where the four velocity of the emitting fluid is $u^\mu$, the photon momentum is $k^\mu$, and $b^\mu=u_\nu \star F^{\nu\mu}$ is the fluid frame magnetic field four-vector, where $\star F^{\mu\nu}$ is the Maxwell tensor. The fluid frame magnetic field $b^\mu$ is easily computed from the lab frame magnetic field $B^i=\star F^{i0}$ in ideal GRMHD using e.g., Eqs. 16-17 in \citet{Gammie2003}. We will assume that the lab frame magnetic field is purely toroidal, so $B^i = (0,0,B^\phi)$.
    
To compute the emissivity $j_\nu$ from \autoref{eq:jemitI}, we need the thermal synchrotron integral,
    \begin{align}
        I_I(x) &= \frac{1}{x}\int_{0}^{\infty} z^2\exp (-z)F \left( \frac{x}{z^2} \right) dz, \label{eq:thermsyncintI}
    \end{align}
where $F(x)$ is the standard synchrotron integral \citep{Rybicki:847173}. We use an analytic approximation to the numerical solution for \autoref{eq:thermsyncintI} in the regime $\nu/\nu_c >1$ given by equation A18 in \cite{Dexter_2016}.
    
Lastly, since we are only modeling thermal electrons, the source function $S_\nu$ is taken to be the Planck black body spectrum function,
    \begin{align}
        S_\nu &= \frac{2h\nu^3}{c^2}\frac{1}{\exp(\frac{h\nu}{2k_BT_e})-1} \label{eq:blackbod},
    \end{align}
where here $h$ is Planck's constant.
    
%___________________________________________________________________________________________________________________________________________________________________________________
      
\subsection{Disk model}
\label{sec:diskModels}

We assume the particles in the disk to move on generic equatorial orbits with a four-velocity~\CPA
\begin{align}
	\label{eq:FourVelocity}
    u=u^t\left(\partial_t-\iota\partial_r+\Omega\partial_\phi\right),
\end{align}
where the angular and radial-infall velocities are defined, respectively, as
\begin{align}
    \label{eq:OmegaAndIota}
    \Omega=\frac{u^\phi}{u^t},\quad
    \iota=-\frac{u^r}{u^t}, 
\end{align}
and $u^\mu$ denote the contravariant components of the four-velocity. In the disk model of~\citet{Cunningham1975}, $\Omega$ is fixed to the Keplerian value and $\iota$ is zero outside the innermost stable circular orbit (ISCO). Inside the ISCO, $\iota$ and $\Omega$ are determined by conserving the energy and angular momentum of particles falling in from the ISCO in the so-called ``plunging region.'' We adopt the parameterized model of~\CTA~which relates $\Omega$ and $\iota$ to their values in the \citet{Cunningham1975} model using three parameters $\left(\xi,\beta_r,\beta_\phi\right)$ which range from 0 to 1. In this model, $\xi$ is the ratio of the fluid angular momentum outside the plunging region to the Keplerian value, and varying $\beta_r$ ($\beta_\phi$) from zero to unity smoothly interpolates the radial (azimuthal) velocity from that corresponding to radial infall from rest at infinity to the sub-Keplerian value. For all the simulations performed in this work we set all three parameters equal to unity, corresponding to the Keplerian disk model~\citep{Cunningham1975}. 

\begin{figure}
        \centering
        \includegraphics[width=7cm]{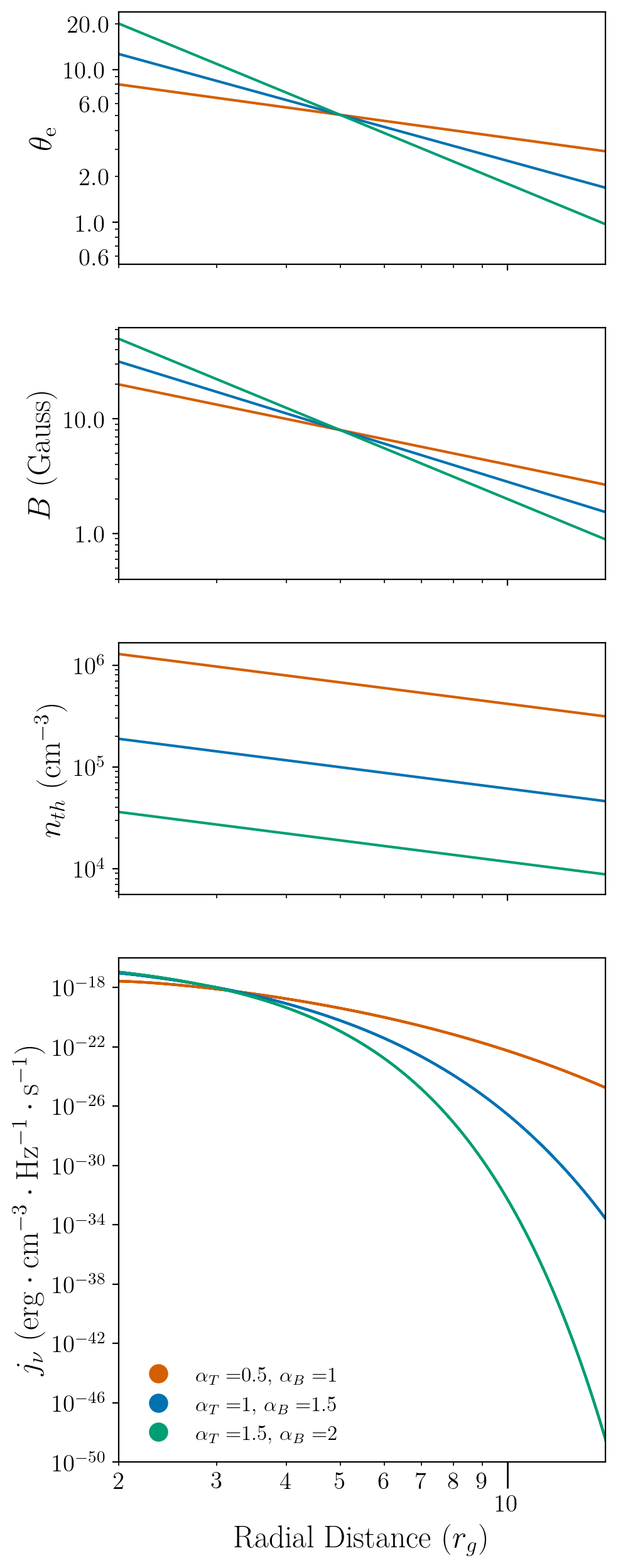}
                \caption{The dimensionless electron temperature, $\Theta_e$ (first), magnetic field strength $B$ (second), particle density $n_{th}$ (third), and emission coefficient $j_\nu$ (fourth) as a function of the radial distance from the center of the black hole for three disk models. The first three quantities are given by \autoref{eq:dens}-\autoref{eq:mag}, respectively. The emission coefficient is calculated using \autoref{eq:jemitI}, and was taken from a radial slice at polar angle $\varphi=\pi$ for an image of a rotating ($a_*=15/16$) black hole, taking $\nu_0=230$ GHz and including the effects of relativistic redshift. The rest of the parameters are fixed and given in \autoref{tab:constModels}. \label{fig:powerlaw}}      
\end{figure}

To specify the density, temperature, and magnetic field in the accretion disk, we use similar power law prescriptions as \cite{2009ApJ...697.1164B}, based on the results of semi-analytic models of hot accretion flows \citep{Yuan2003}:
\begin{align}
        n_{th} &= n_{th,0}(r/R_b)^{-\alpha_{n}}, \label{eq:dens}\\
        T_e &= T_{e,0}(r/R_b)^{-\alpha_{T}}, \label{eq:temp} \\
        B &= B_0(r/R_b)^{-\alpha_{B}}.  \label{eq:mag}
\end{align} 
Our model thus has the following important parameters: $n_{th,0}$, $T_{e,0}$, $B_0$, which define the value of their respective functions at the radius $R_b$; we fix $R_b= 5 {r_{\rm g}}$ as the value used in simple one-zone models of M87 using the EHT results \citep{EHTM87I}. The power law indices $\alpha_n$, $\alpha_T$, and $\alpha_B$ are all defined to be positive. In~\autoref{tab:constModels} we provide a summary of all the model parameters, their ranges and fiducial values, and in \autoref{fig:powerlaw} we provide examples of the profiles used in our model.

\begin{deluxetable*}{l l c}
        \tabletypesize{\scriptsize}
        \tablewidth{\textwidth}
        \tablecaption{Model Parameters}
        \tablehead{
            \colhead{Parameter}
            & \colhead{Description}
            & \colhead{Value(s)}
            }
        \startdata
            $M_{\rm BH}$ & Black hole mass{$^\dagger$} & $6.5 \times 10^9 M_\odot$ \\
            $\theta_{\rm o}$ & Observer's inclination angle{$^\dagger$} & $17 \degree$ \\ 
            $a_*$ & Black hole spin & $[0.0,0.5,15/16]$\\
            $\nu_0$ & Observation frequency & $10$ - $670$ GHz \\ 
            $\theta_{\rm disk}$ & Disk opening angle{$^\dagger$} $h/r$ & $0.5$ \\
            $R_b$ & Distance at which the power laws take base values{$^\dagger$} & $5{r_{\rm g}}$\\
            $T_{e,0}$ & Base value for the temperature power law & [$3$, $5$, $7$]$\times 10^{10}$K \\
            $B_0$ & Base value for the magnetic field power law{$^\dagger$} & $8$ Gauss  \\
            $c_r$ & Magnetic field radial component coefficient{$^\dagger$} & $0$ \\
            $c_\phi$ & Magnetic field toroidal component coefficient{$^\dagger$} & $1$ \\
            $\alpha_{n}$ & Sets decay speed for the density power law{$^\dagger$} & $0.7$ \\
            $\alpha_T$ & Sets decay speed for the temperature power law & $[1.5,1,0.5]$\\
            $\alpha_{B}$ & Sets decay speed for the magnetic field power law & $[2,1.5,1]$\\
        \enddata

        \tablecomments{Summary of the thirteen parameters for the black hole and accretion disk. The mass and spin of the black hole and the observer's inclination are entirely geometrical (i.e., independent of the astrophysics of the disk). Parameters with values in brackets were varied, while those with a dagger were kept constant. All combinations of varied parameters were used in individual models. For most of the examples in this work, unless otherwise stated, we set $T_{e,0}=3 \times 10 ^ {10}$~K, as this value produces a flux peak around $\nu_0=230$ GHz comparable to that of M87* for the fiducial model $\alpha_T=1.0$, $\alpha_B=1.5$, and $a_*=15/16$. The density normalization, $n_{th,0}$, for each model is set such that the corresponding model has a total flux density of $\sim0.5 \rm Jy$, as observed in M87*~\citep{EHTM87V}. Details on the normalization procedure are presented in \autoref{appendix:norm}.}
        \label{tab:constModels}
\end{deluxetable*}

We fix the black hole mass and distance to that of M87*, $M_{\rm BH}=6.5 \times 10^9 M_\odot$ \citep{EHTM87I} and $D=16.8$ Mpc~\citep{EHTM87VI}, respectively. To allow for direct comparisons between models, we determine the density scale $n_{th,0}$ such that the total flux density for the full RTE solution image at $\nu_0=230$ GHz is $F_{230}=0.5$~Jy, roughly the same as observed for M87* \citep{EHTM87IV}. This density normalization procedure is also standard when comparing GRMHD simulations to EHT data~\citep{EHTM87V,EHTM87VIII}. Following~\citet{2009ApJ...697.1164B}'s fitting of M87* radio spectral index, we fix the fiducial value for the density power law to $\alpha_n=0.7$. 
Different values for the exponent in the functional form of the magnetic power law allow us to smoothly transition between the falloff expected for purely toroidal (e.g., SANE) dominated field lines ($B\propto 1/r$) and poloidal (e.g., MAD) field lines ($B \propto 1/r^2$). 

The prescription for the intensity described in the previous sections does not include a black hole jet or any emission from outside the equatorial plane. Although there has been debate over the location of the peak of the emission, whether it be from the disk or jet, there is an extensive past history of focusing near-horizon emission in the disk \citep{2009ApJ...697.1164B}. Nevertheless, equatorial models without jet emission can produce similar images to near horizon GRMHD simulation images \citep{JMD2020,Cardenas-Avendano:2022csp}. Near the event horizon, a wide opening angle jet (as expected of M87*) will have its region of peak emission near the equatorial plane \citep{EHTM87I}. Thus, the lack of explicit jet emission should not limit our model being used for near-horizon studies of M87*. 
% ___________________________________________________________________________________________________________________________________________________________________________________
 
% __________________________________________________________________________________________________________________________

\section{Time-averaged simulated black hole images} 
\label{sec:floats}

As summarized in~\autoref{tab:constModels},we now have a large parameter space to explore. We now produce and analyze black hole images and, for each model, we calculate their SED, average ring radii, and optical depth, broken down by the $n$ sub-images. For the majority of the analysis, we fix the electron temperature at $r=R_b$ and focus on the $T_{e,0}=3 \times 10^{10}$~K model subspace. In \autoref{sec:spectra} we investigate the effects of changing the temperature power normalization, $T_{e,0}$.

\subsection{Image Production}
\label{sec:Images}

For each point in the parameter space specified in \autoref{tab:constModels}, we generate images for observation frequencies $\nu_0$ ranging from $10$ to $670$ GHz at intervals of $20$ GHz, resulting in a total of $34$ images per model. All images generated in this work were computed in a square format with dimensions of $[-25\,r_{\rm g},25\,r_{\rm g}]$ and a resolution of $8000\times8000$ pixels, i.e., a pixel size of $0.00625\,r_{\rm g}$. This is a conservative resolution chosen to obtain accurate radii measurements across all the models and frequencies studied (see~\autoref{appendix:convergence} for further details). In \autoref{fig:FullImages} we show images at the frequencies $\nu_0=$$[90, 230, 350]$ GHz for models with spin $a_*=15/16$, $\alpha_B=1.5$, and $T_{e,0}=3 \times 10^{10}$~K, as we vary the temperature power law index $\alpha_T$. 

\begin{figure*}[hb!]
    \plotone{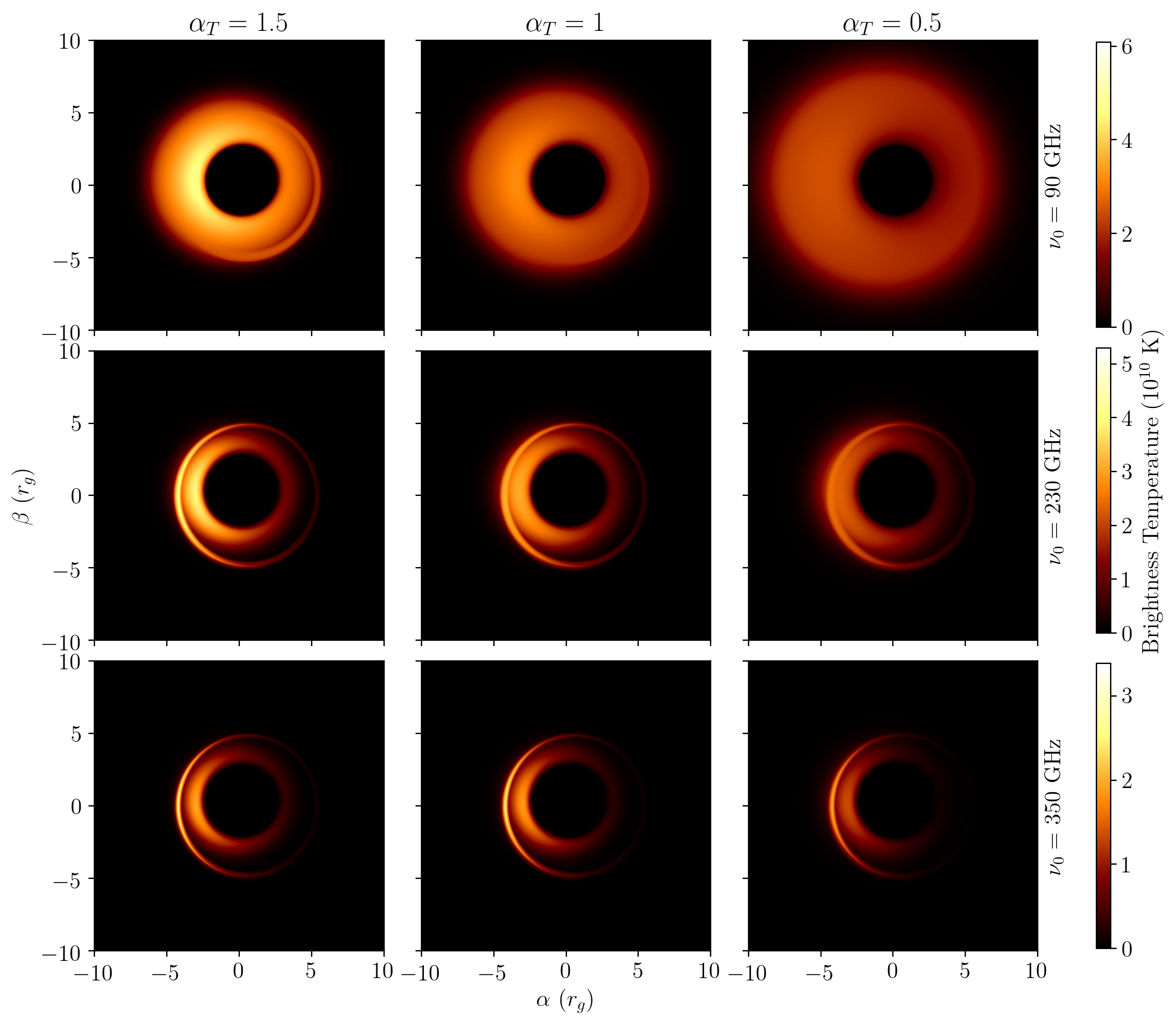}
    \vspace{12pt}
    \caption{Images produced for different temperature power law indices $\alpha_T$, and observation frequencies. Columns correspond to different $\alpha_T$ values, while rows correspond to changes in the frequency of observation ($\nu_0$). The middle row's observation frequency is taken as the $230$ GHz measured by the current EHT, while the top (90 GHz) and bottom (350 GHz) row values were chosen for similarity to the upcoming advances in EHT observation frequencies \citep{Chael_Issaoun_Pesce_Johnson_Ricarte_Fromm_Mizuno_2023}. Other model parameters were fixed to $a_*=15/16$, $\alpha_B=1.5$, and $T_{e,0}=3 \times 10^{10}$~K. The remaining parameters are shown in \autoref{tab:constModels}. Images across each row share the same color bar. Note the large variance in the shape of the $n=0$ images, in strong contrast to the higher $n$ rings, which are less sensitive to the astrophysical parameters. Also note that the higher $n$ rings become more prominent in the image for higher $\nu_0$ (as the flow becomes optically thin) and larger $\alpha_T$. \label{fig:FullImages}}
\end{figure*}

The top row of images in \autoref{fig:FullImages} at $\nu_0=90$ GHz can be easily distinguished from the rest by the large dull halo, the $n=0$ sub-image. At low frequencies, this ring is, as expected, optically thick and dominates the image while obscuring the thinner, highly lensed $n=1$ and $n=2$ rings. Within the $\nu_0=90$ GHz row, a larger value of $\alpha_T$ corresponds to a smaller halo, as the emission is more concentrated close to the black hole.  

We find that for the $n=0$ emission, changes in the temperature index $\alpha_T$ have a larger impact on the image structure than changes in the magnetic field index $\alpha_B$. This is expected, as at lower frequencies the image is marginally optically thick, where the thermal source term dominates. At higher frequencies, all astrophysical parameters become important as the lower optical depth permits emitted synchrotron radiation to travel further before re-absorption.

At higher frequencies the higher-order $n$ photon rings in \autoref{fig:FullImages} are easier to distinguish. These higher-order rings have approximately the same size and shape for all values of $\alpha_T$, as the range of sub-image radii is more highly determined by geometrical parameters like the black hole spin observer's inclination angle, than by the properties of the plasma, such as $\alpha_T$.\footnote{Higher-order rings will be (exponentially) demagnified, rotated, and time-delayed images of the previous contribution~\cite{GrallaLupsasca2020p2}. The exponential demagnification can be as little as~$10\%$ for the first photon ring (for rapidly rotating black holes observed at a high inclination angle), increasing significantly with higher photon rings.} In general, higher frequency and more highly-lensed images are expected to be less dependent on the precise values of the astrophysical parameters such as $\alpha_B$ and $\alpha_T$ \citep[e.g.,][]{JMD2020,Chael_Johnson_Lupsasca_2021,Cardenas-Avendano:2022csp}.

\begin{figure*}[ht!]
        \plotone{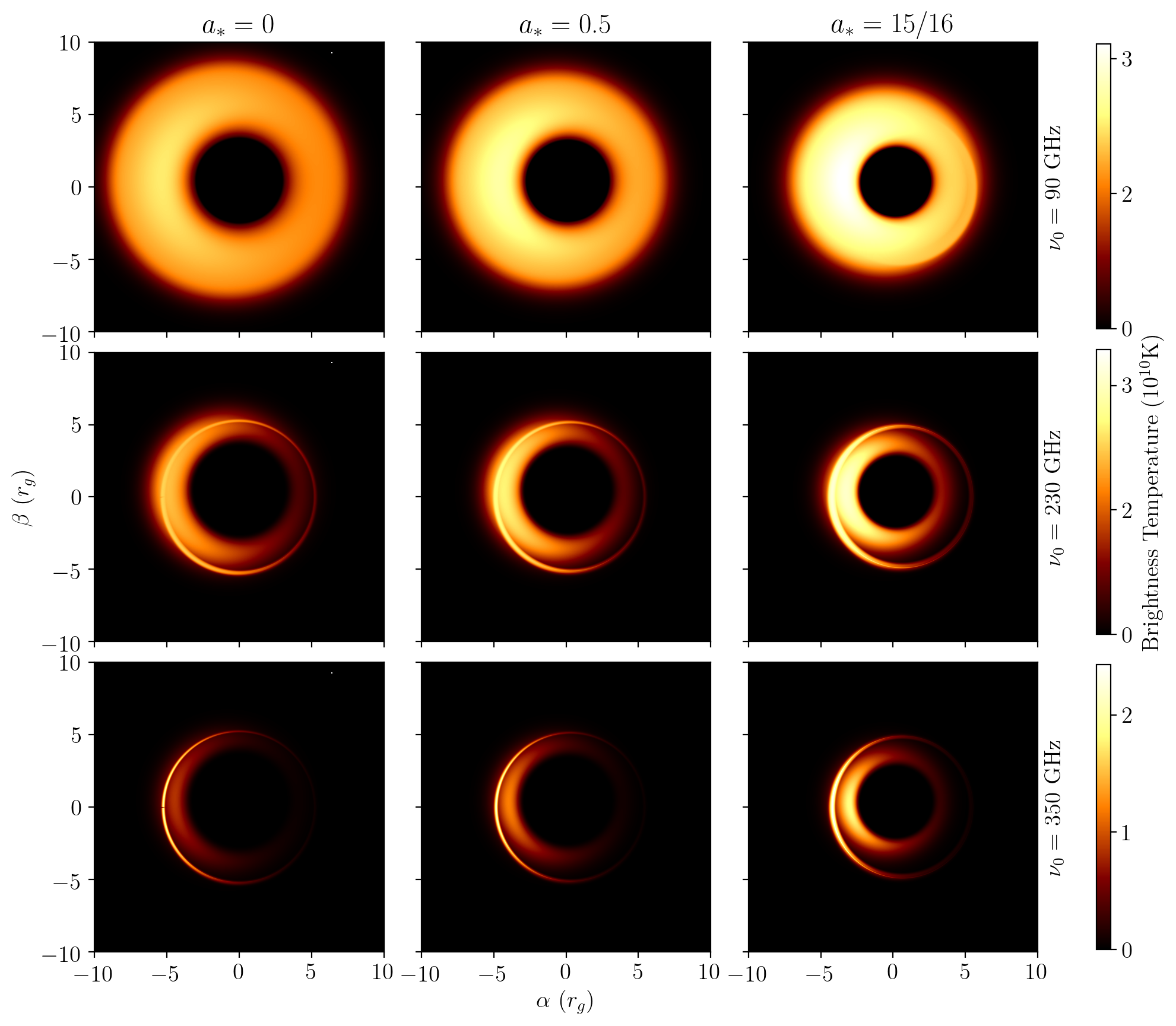}
        \vspace{12pt}
        \caption{Images produced for different values of the black hole spin and observation frequency. From left to right, the spin increases ($a_*=0,0.5,15/16$), while from top to bottom, the observation frequency is increased ($90,230,350$~GHz). For all of these images, $\alpha_T=1$, $\alpha_B=1.5$, and $T_{e,0}=3 \times 10^{10}$~K. The remaining parameters are fixed to the fiducial values presented in \autoref{tab:constModels}. Note the dependence of the size of the ``inner shadow'' on spin $a_*$~\citep{Chael_Johnson_Lupsasca_2021}. 
        \label{fig:SpinFullImages}}
\end{figure*}

To examine the image dependence on the spacetime parameters, \autoref{fig:SpinFullImages} displays images for $\alpha_T=1$ and $\alpha_B=1.5$ at different spin values and observation frequencies. In contrast to \autoref{fig:FullImages}, in these images the $n=0$ halo varies much less across each row. When changing the spin $a_*$ but fixing the astrophysical parameters the size of the $n=0$ emission notably shrinks as the inner edge of the emitting region moves closer to the black hole. This inner edge (known as the ``inner shadow'': \citealt{Chael_Johnson_Lupsasca_2021}) of the $n=0$ image is strongly dependent on the spin value $a_*$ and is independent of $\nu_0$. The size of the $n=1$ and $n=2$ rings also shrink with spin \citep{JohannsenPsaltis}, further exemplifying the sensitivity of the higher-order subimages to black hole parameters. 

\subsection{Inferred Ring Sizes}
\label{sec:ringSizes}
    
We now define a ring radius, $\rho(\varphi)$, for each image at a given position angle $\varphi$. This process involves performing a linear interpolation of the underlying image, converting the Bardeen coordinates, $(\alpha,\beta)$, to polar coordinates $(\rho', \varphi)$ and calculating the intensity $I_\varphi(\rho')$ on radial slices from the origin of the image plane $\rho'_{\rm min}=0$ to $\rho'_{\rm max}=10 r_{\rm g}$. We define the ring radius $\rho$ at angle $\varphi$ to be the radius at which the radial intensity slice peaks, i.e., $\rho(\varphi)=\mathrm{argmax}\left[I_\varphi(\rho)\right]$. 

\begin{figure*}[ht!]
        \centering
        \plotone{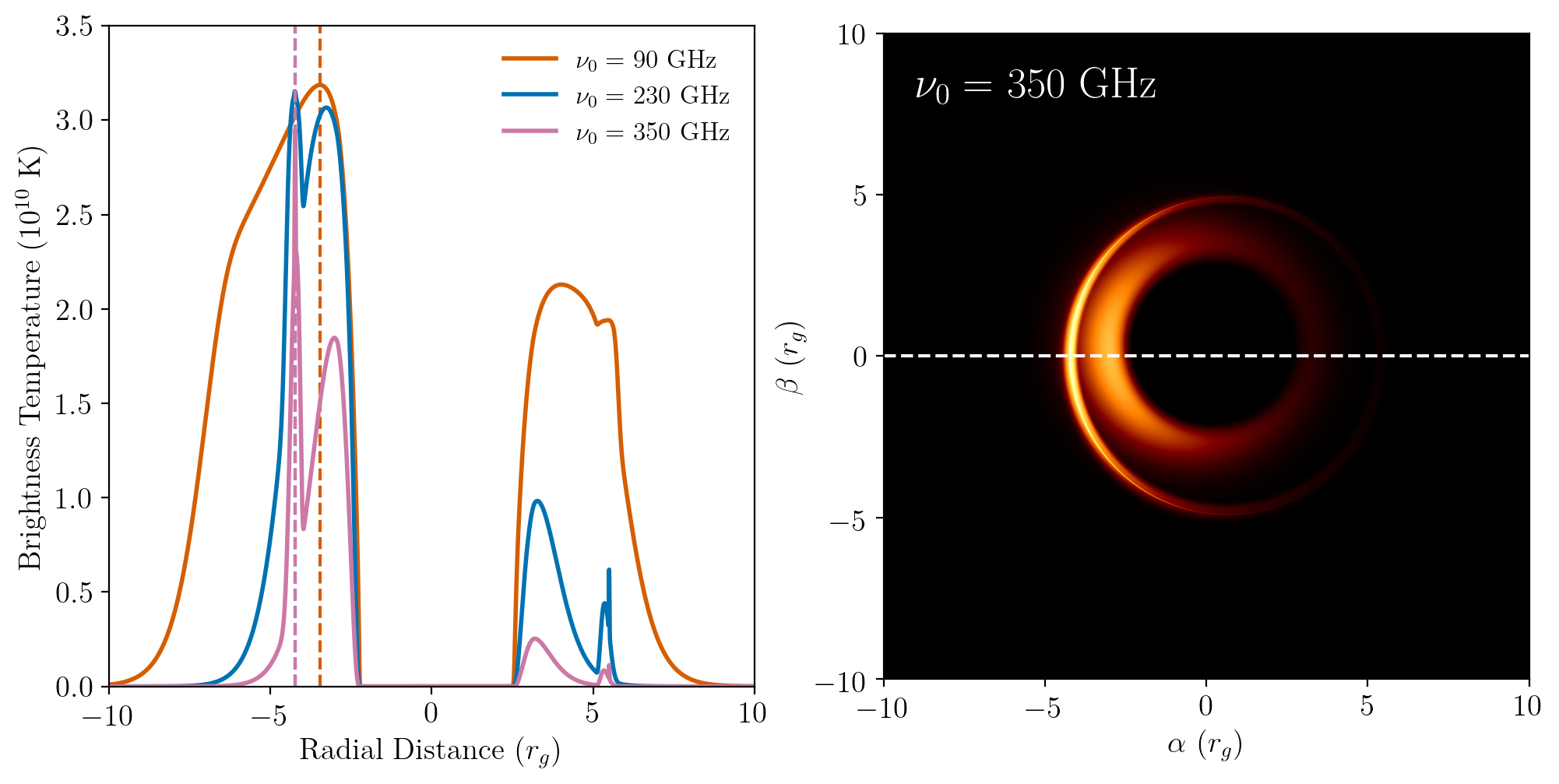}
        \caption{(Left) The brightness temperature as a function of the radial distance (measured in $r_{\rm g}$) on the horizontal axis for a model observed at $\nu_0=90,230,350$~GHz in orange, blue, and pink, respectively. The vertical lines indicate the intensity peak, which, as demonstrated in these examples, shifts with frequency. (Right) A black hole image at $\nu_0=350$~GHz featuring a horizontal radial slice corresponding to the pink brightness profile on the left panel. For this example we have set $a_*=15/16$, $\alpha_T=1$, $\alpha_B=1.5$, and $T_{e,0}=3 \times 10^{10}$ K. The remaining parameter values are given in \autoref{tab:constModels}. Note that as the image changes from the left side ($\varphi=\pi$) to the right ($\varphi=0$), the overall brightness temperature decreases across all rings, but the distance in the peak brightness points between the $n=0$ and $n=1,2$ rings increases. 
        }
        \label{fig:radiiCalc}
    \end{figure*}

In \autoref{fig:radiiCalc}, we show radial profiles $I_\varphi(\rho')$ for a high spin $a_*=15/16$ model at different frequencies for two position angle cuts, $\varphi=0$ (right side of the image) and $\varphi=\pi$ (left side of the image). Both the redshift factor $g$ in \autoref{eq:invarientJ} and the disk power laws in \autoref{eq:jemitI} play a role in the resulting image structure. The rest frame emissivity determined by the disk power laws increases toward the horizon while the redshift factor $g$ decreases toward the horizon from gravitational redshift. As the image is dimmer on the right side, the redshift $g$ factor is stronger on this side of the image from the combined effects of gravitational redshift and the special relativistic (Doppler) redshift from the receding disk velocity along the line of sight. Lower values of $g$ push the intensity peak out to larger radii on the right side of the image. The combined gravitational and special relativistic redshift also causes an overall brightness asymmetry in the image; all rings are
less bright on the disk's receding (right) side than the approaching (left) side. 

    \begin{figure}[h!]
        \plotone{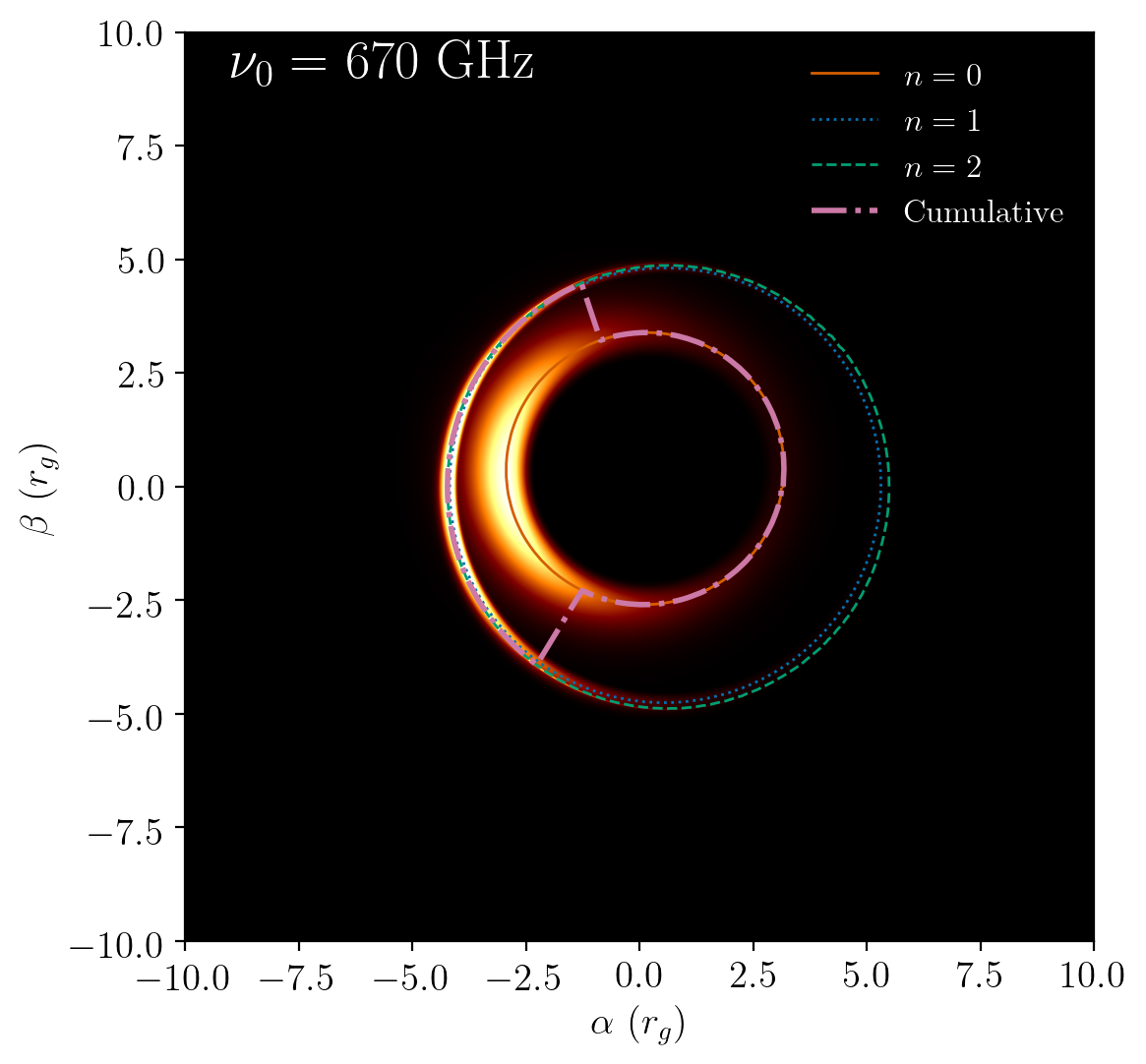}
        \caption{Ring radii curves $\rho(\varphi)$ measured on an image of the fiducial model ($a_*=15/16$, $\alpha_T=1$, $\alpha_B=1.5$, and $T_{e,0}=3 \times 10^{10}$ K) at 670 GHz. The remaining model parameters can be seen in \autoref{tab:constModels}. There is a ``jump'' of the overall brightness peak radius from the $n=2$ ring (green-dashed line) to the much smaller $n=0$ (orange solid line) sub-image between the left and right sides of the image.
        \label{fig:670GHZ}}
    \end{figure}
        
The difference in radius between the left and right sides of \autoref{fig:radiiCalc} increases for higher-order photon rings, i.e., the position of the the intensity peaks in each subimage shifts right relative to the $\alpha=\beta=0$ origin. This effect is the result of the asymmetries the spin induces in the strong gravitational lensing. As a result, the photon rings shift relative to the origin \citep[e.g.,][]{JohannsenPsaltis} more than the direct emission. 

These asymmetries lead to another observable effect: the cumulative image's peak brightness point does not always trace a single photon ring for all frequencies and image slices. For instance, the cumulative image's peak corresponds to the peak of the $n=0$ subimage for the three frequencies shown in~\autoref{fig:radiiCalc} for $\varphi=0$ (right-hand side of the image). However, for $\varphi=\pi$ (left-hand side of the image) the cumulative image's peak corresponds to the peak of the $n=2$ subimage for the $\nu_0=$ $230$ GHz and $350$ GHz images on the left side of the image. 
Changes in which subimage contributes the peak brightness point around the ring is again tied to the geometric asymmetry induced by the disk's motion and the black hole's spin. On the right of the image, higher $n$ rings that would otherwise contribute to the peak brightness value are dampened by redshift, while the brighter $n=0$ sub-image is concentrated to smaller radii because of the choice of the disk parameters. Thus, for the model shown in~\autoref{fig:radiiCalc} the offset between the $n=0$ and $n=1,2$ radii does not allow for the characteristic ``wedding cake'' structure, which stacks the higher order photon rings on top of the direct emission to exceed the peak brightness of the direct image \citep{JMD2020}. On the left side of the image, the larger relative blueshift of the higher photon rings does allow their brightness to exceed that of the direct image, $n=0$.  

The overall result of these asymmetries can be seen in \autoref{fig:670GHZ}, where we plot $\rho(\varphi)$ for the full $670$~GHz image and for the individual $n=0,1,2$ subimages. The large spin value in this example causes the higher order rings on the right side of the image to be much dimmer than the $n=0$ sub-image; as a result, the overall brightness peak in the image jumps from $n=2$ on the left to $n=0$ on the right. These effects vanish for non-spinning black holes where the brightness asymmetry is reduced and the cumulative ring radii remain attached to $n=2$ for all $\varphi$.

The $n=0$ direct image radius in the optically thin regime is set by the radius where the emitted frequency is approximately equal to the critical frequency, $\nu \approx \nu_c$. As $\nu$ increases, since $\nu_c \propto B\Theta _e^2$, this rough equality shifts closer to the horizon. The radius of the higher order $n$ rings are instead dominated by the position of the critical curve, set by the black hole mass and spin. 

\subsubsection{Mean Radii across Observation Frequencies}
    
To make radius comparisons across different observer frequencies $\nu_0$, we average the radial function $\rho(\varphi)$ into a single value, $\bar{\rho}$, for each image. To compute this average we use the method detailed in Appendix B of~\cite{Chael_Johnson_Lupsasca_2021}. In short, we compute the second moment (covariance) matrix $\Sigma$ for the shape $\rho(\varphi)$, diagonalize it, and find the principal axes $a$ and $b$ of the shape $\rho(\varphi)$. These define the average radius by the quadrature mean $\bar{\rho} = \sqrt{(a^2 + b^2)/2}$. 

\begin{figure*}[h!]
        \plotone{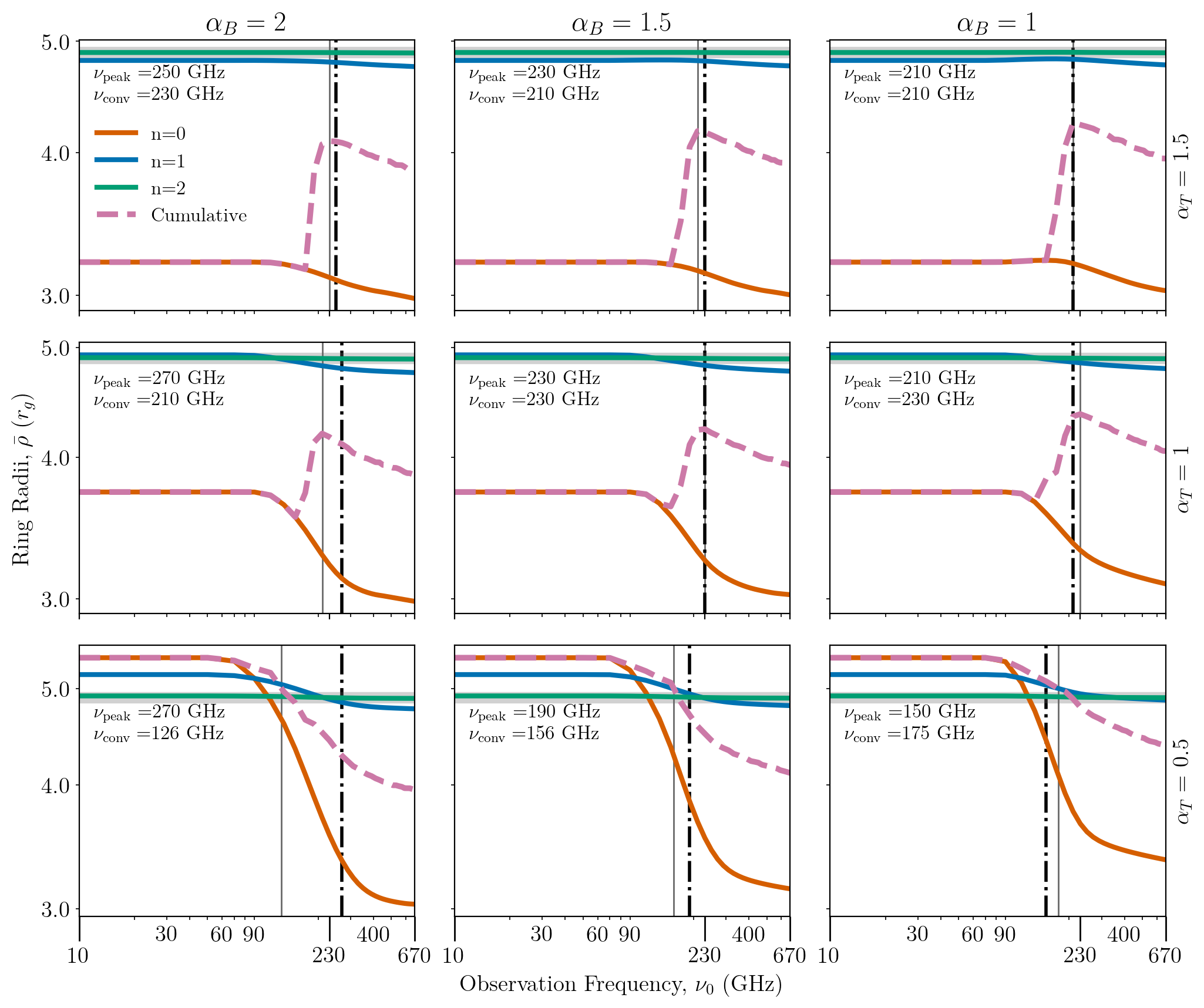}
        \caption{Mean ring radii $\bar{\rho}$ (measured in $r_g$) as a function of the observation frequency, for different values of $\alpha_T$ and $\alpha_B$. For all of these examples, we have set to $a_*=15/16$ and $T_{e,0}=3 \times 10^{10}$~K, and all other parameters as shown in \autoref{tab:constModels}. The results for the $n=0$ image and the individual photon ring contributions $n=1,2$ are shown in orange, blue, and green, respectively; the radii measured from the full image including contributions from all photon rings are shown in purple. The thin gray vertical line indicates the value of $\nu_{\rm conv}$, which is the frequency that meets the criterion we have defined for the transition to the optically thin regime. The dashed-dotted vertical line indicates $\nu_{\rm peak}$, i.e., the frequency $\nu_0$ at which the total flux density is at a maximum. The grey horizontal line is the average radius of the critical curve, or black hole shadow. At low observation frequencies in the optically thick regime, the choice of temperature power law $\alpha_T$ uniquely determines the ring radii. As the flow becomes optically thin, the magnetic field power law index $\alpha_B$ produces a measurable effect on the ring radius. 
        }\label{fig:ThickRadVNuGrid}
    \end{figure*}

In \autoref{fig:ThickRadVNuGrid}, we show the average radius $\bar{\rho}$ as a function of frequency $\nu_0$ for a model with $a_*=15/16$ and  $T_{e,0}=3 \times 10^{10}$ for different values of $\alpha_T$ and $\alpha_B$. In this figure, we compute the average radii $\bar{\rho}$ for both the total image and for the individual subimages (using \autoref{eq:simpleAbsorb}). The radius that changes the most when changing the parameters of the model is the measurement of the $n=0$ sub-image. As seen in \autoref{fig:photonRings} and discussed previously, the higher valued $n$ rings are defined by spending a longer path length in close proximity to the black hole's horizon. As such, the higher $n$ sub-images are constrained to a narrow range of radii determined by the black hole and observer parameters~\citep{GrallaLupsascaMarrone2020,Cardenas-Avendano:2023dzo}. This effect is seen by the smaller changes in the $n=1$ ring radius with frequency when compared to the $n=0$ radius. The $n=2$ ring closely follows the radius of the critical curve or ``black hole shadow'' radius.

In all panels of \autoref{fig:ThickRadVNuGrid}, the total or cumulative image has a radius identical to that of the $n=0$ direct image up to a certain transition frequency. This occurs because the flow is optically thick at low frequencies, and the higher order sub-images are not visible. Below this transition frequency, the image radius is constant with frequency, as the apparent image size is set purely by where the redshifted temperature peaks in the optically thick limit. At this transition frequency, the cumulative image begins to see contributions from the $n=1$ and $n=2$ sub-images.

To mark the transition frequency at which the disk becomes optically thin and allows the lensed sub-images to be seen, we define a convergence frequency $\nu_{\rm conv}$, and plot it in \autoref{fig:ThickRadVNuGrid} with vertical grey lines. We define this frequency as either the frequency where the cumulative image radius $\bar{\rho}$ comes within $2\%$ of the $n=2$ sub-image radius (Type I), or, if $\bar{\rho}$ never reaches that threshold, as the frequency at which the difference between the cumulative and the $n=0$ radii is at a minimum (Type II). The type of transition frequency $\nu_{\rm conv}$ a model experiences strongly depends on the choice of the temperature power law $\alpha_T$ and the choice of the electron temperature normalization at $r=5r_{\rm g}$, $T_{e,0}$, as we explore in more detail in \autoref{sec:spectra}. 

At frequencies much lower than $\nu_{\rm conv}$, the optical depth is large enough to create a ring ``surface'' below which no details of higher $n$ rings can be resolved. At high frequencies, emission from the higher $n$ rings can be seen, allowing for the brightness peak location of the cumulative image to be noticeably different from that of the individual rings. As discussed in \autoref{sec:ringSizes}, when the $n=0$ emission is at large radii comparable to the shadow radius, the $n=1$ and $n=2$ emissions are superposed or stacked on top of the optically thin $n=0$ sub-image, creating a ``wedding-cake'' structure so that the $n=2$ ring becomes the brightest point in the image. However, when the $n=0$ direct image is concentrated close to the black hole, these contributions become disconnected from the $n=0$ emission, and the $n=0$ sub-image can remain the brightest point in the image at certain position angles. In these cases, the cumulative image radius $\bar{\rho}$ will remain close to the $n=0$ value, or lie between the radius of the $n=0$ and $n=1$ images, depending on which fraction of the image position angles have their brightest points set by $n=0$ and which are dominated by higher-order sub-images.  

    \begin{figure*}[ht!]
        % \plotone{Figures/SpinThickRadiiVNuRow.png}
        \centering
        \includegraphics[width=18cm]{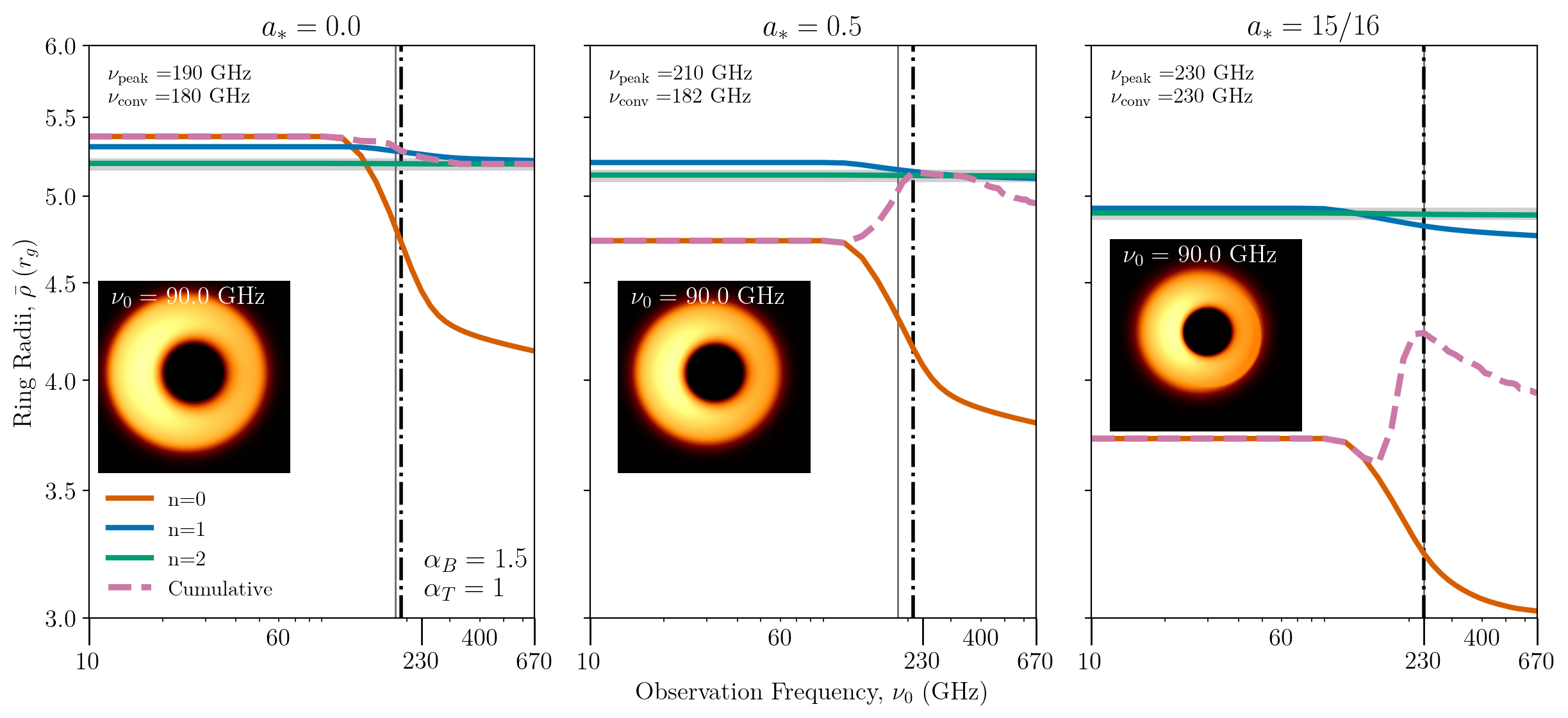}
        \caption{Mean ring radii $\bar{\rho}$ (measured in units of $r_{\rm g}$) as a function of observation frequency for three different values of the black hole spin. The inset images display the respective model produced at $\nu_0=230$ GHz. The dash-dotted black vertical line indicates where the flux density peaks. $\nu_{\rm peak}$.  The thin vertical solid lines indicate the value of $\nu_{\rm conv}$, which marks the transition to the optically thin regime, as defined in~\autoref{sec:ringSizes}. For these examples we have set $\alpha_T=1$, $\alpha_B=1.5$, $T_{e,0}=3 \times 10^{10}$~K and the remaining model parameters are chosen as shown in \autoref{tab:constModels}. The low opacity horizontal grey line under the $n=2$ radii is the critical curve radius.\label{fig:SpinThickRadVNuRow}}
    \end{figure*}

    \begin{figure*}[ht]
        \includegraphics[width=16cm]{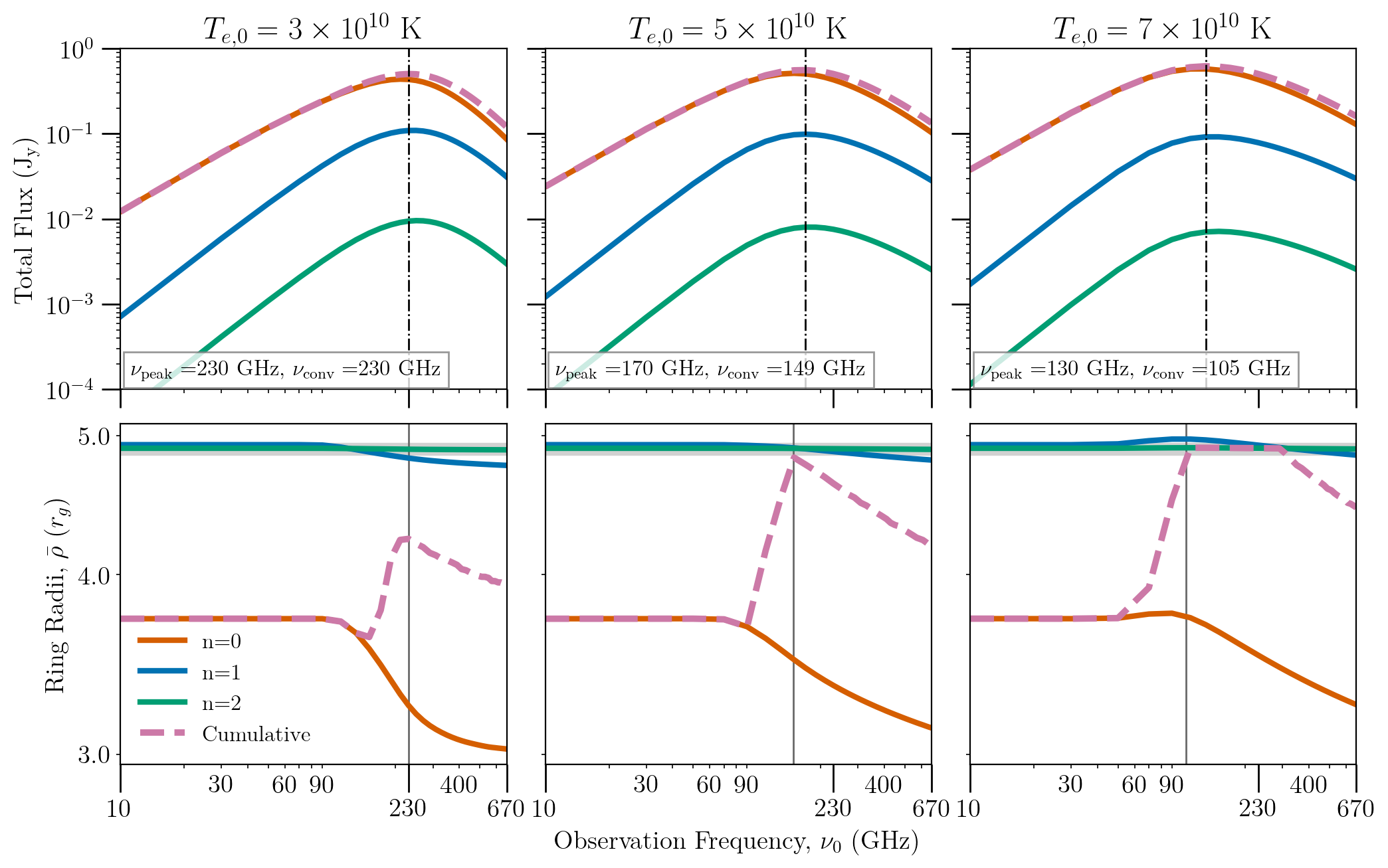}
        \centering
        \caption{Total flux density (top) and averaged mean ring radius (bottom) as a function of observation frequency $\nu_0$ for different $T_{e,0}$ models, with $\alpha_B=1.5$, $\alpha_T=1$, and $a_*=15/16$. All other parameters are set with the fiducial values shown in~\autoref{tab:constModels}. Dashed lines indicate the optically thin assumed RTE solutions. The dash dotted lines indicate $\nu_{\rm peak}$, the frequency at which the total flux density is maximized, while the thin vertical lines indicate $\nu_{\rm conv}$, the frequency that marks the transition to the optically thin regime, as defined in~\autoref{sec:ringSizes}.}
        \label{fig:te0Models}
    \end{figure*}

For the spin $a_*=15/16$, the only models that are type I and have a cumulative average radius that crosses the critical curve radius have a shallow temperature index $\alpha_T=0.5$. This softer falloff of temperature with radius allows the $n=0$ sub-image to be larger than the radius of the $n=2$ ring at low frequencies $\nu_0$. As the $n=0$ direct image becomes optically thin, it begins to shrink; the total image radius contains contributions from both $n=0$ and $n=1,2$, so it also drops with frequency.
In this way, the cumulative ring radius drops in tandem with the radius of the $n=0$ sub-image radius, but offset by a few $r_{\rm g}$. At higher values of $T_{e,0}$, however, some models with small radii at low frequencies (higher $\alpha_T$) do have radii that jump to the $n=1,2$ radius as the $n=0$ direct image becomes optically thin. Those models are discussed in \autoref{sec:spectra}.

As noted before, the choice of temperature power law determines the ring radius at low frequencies where the disk is optically thick. This effect can also be seen in \autoref{fig:ThickRadVNuGrid} by the increase of the low-frequency image radius with decreasing $\alpha_T$ and the insensitivity of the low-frequency radius to $\alpha_B$. These radii range from $\sim3.2r_{\rm g}$ for $\alpha_T=1.5$, to $\sim5.2r_{\rm g}$ for $\alpha_T=0.5$. This behavior is expected, since a face-on optically thick disk will have its peak brightness at a radius where the redshifted temperature is maximized. Disks with steeper temperature profiles (larger $\alpha_T$) will have their redshifted temperature maximized at smaller image radii compared to disks with shallower profiles (smaller $\alpha_T$). 

Another notable trend is the tendency of $\nu_{\rm conv}$ to be on either side of the point where the total flux density SED peaks, $\nu_{\rm peak}$ (which we indicate with a dashed vertical line in \autoref{fig:ThickRadVNuGrid}). Large values of $\alpha_T$ and small values of $\alpha_B$ correspond to $\nu_{\rm conv}$ ocurring after $\nu_{\rm peak}$, while small values of $\alpha_T$ and large values of $\alpha_B$ models have $\nu_{\rm conv}$ before $\nu_{\rm peak}$. This trend is most clearly seen in the high spin $a_*=15/16$ model. Regardless, in all cases but one in~\autoref{fig:ThickRadVNuGrid}, the transition radius $\nu_{\rm conv}$ measured in the image domain is within $\sim50$GHz of the peak SED frequency $\nu_{\rm peak}$, suggesting that the peak SED frequency, $\nu_{\rm peak}$, is a good estimator of the frequency at which the higher order photon rings begin to contribute to the image structure. 

The transition frequency $\nu_{\rm conv}$ is a measurement made only from the image as a function of frequency (without any knowledge of the underlying disk parameters) that marks the transition from the optically thin to thick regime. It must be noted that the actual transition does not occur isotropically across the image, as the redshift factor and thus the emitted frequency depends upon the direction of accretion flow relative to the observer~\citep{Broderick_Fish_Doeleman_Loeb_2011}. The $\nu_{\rm conv}$ metric does not capture this complexity, as it is an average over the entire image at each frequency point. The complexities mentioned emphasize the importance of being careful when defining a single ring radius metric for black hole images in the image domain. At most high frequencies in our models, the cumulative ring radius contains contributions from multiple photon rings, rather than tracking $n=0$ or $n=1,2$ exactly.

\subsubsection{Mean Radius Dependency on the Black Hole's Spin}

To analyze the dependence of the image radius on spin and frequency, \autoref{fig:SpinThickRadVNuRow} shows results from the fiducial model parameters $\alpha_T=1$ and $\alpha_B=1.5$ at different black hole spins $a_*=[0,0.5,15/16]$. We see that, as expected, the $n=2$ and $n=1$ radii decrease with spin, as they closely track the black hole's critical curve.
The $n=0$ radius also shrinks with spin with the other parameters held constant, as the emission region extends closer to $r=0$ as the projected black hole horizon shrinks. The difference between the radius convergence frequency $\nu_{\rm conv}$ and the SED peak frequency $\nu_{\rm peak}$ does not vary significantly over this spin range, having at most a difference of $21$ GHz. However, at higher spin the flow remains optically thick to higher frequencies, with an increase of $\sim 55$ GHz from $a_*=0$ to $a_*=15/16.$

Most interesting is the change in the relative position for the $n=0$, $n=1$ and $n=2$ radii. Lower spins correspond to larger radii for all rings. In the $a_*=0$ model, the $n=0$ sub-image radius is larger than the critical curve for low frequencies; for $a_*=0.5$ and $a_*=15/16$, it is smaller. As a result, in the $a_*=0$ model the cumulative image radius only shrinks with frequency and approaches the $n=2$ value; in $a_*=0.5$ and $a_*=15/16$, the total image radius increases as contributions from the higher order rings become important at low optical depth. Furthermore, the $a_*=0$ model shows much smaller asymmetry across the image from relativistic red/blueshift -- as a result, the $n=1,2$ rings remain the brightest points in the high frequency $a_*=0$ images at all position angles, and the cumulative image radius at high frequencies is approximately the $n=2$ radius. In the high spin models, the asymmetry induced by the black hole's spin causes the $n=1,2$ rings to be less bright than the direct image on the receding side of the flow; as a result, the cumulative radius lies in between the individual $n=0$ sub-image radius and $n=1,2$ rings radii. This suggests that the predominant source of image asymmetry in these models is sourced by the black hole spin, and not the Keplerian rotation \citep{EHTM87V}.
    
\subsection{Model SEDs and Disk Temperature Dependence \label{sec:spectra}}

We now compute the total flux density of each image in Jansky (Jy). In \autoref{fig:te0Models} we show SEDs for three models. As expected, for all three cases the $n=0$ emission dominates the total flux density. Since the spectra in the $\alpha_B$-$\alpha_T$ plane have relatively little variation, we focus on comparing models with different $T_{e,0}$ values for fixed $\alpha_B=1.5$, $\alpha_T=1$.
We find that, as expected, increasing $T_{e,0}$ shifts the spectral peak $\nu_{\rm peak}$ to lower frequencies; it also shifts the image domain transition $\nu_{\rm conv}$ in tandem with $\nu_{\rm peak}$. Models with higher $T_{e,0}$, for example, had all $\nu_{\rm conv}$ after $\nu_{\rm peak}$, meaning that $\nu_{\rm peak}$ itself may be used as a conservative estimate of the transition wavelength from optically thick to thin emission.

Interestingly, the hotter models have a larger range of frequencies where the cumulative image has its brightness maximum entirely set by the higher order photon rings (as shown in the bottom row of \autoref{fig:te0Models}). This is because for the hotter models, the $n=0$ direct image radius does not decay as rapidly with frequency after the flow becomes optically thin. For a range of frequencies, then, the $n=0$ sub-image is sufficiently bright at larger radii to elevate the $n=1,2$ rings to the brightest point in the image in the ``wedding cake'' structure \citep{JMD2020}. In cooler models, the $n=0$ sub-image radius decays more rapidly and it becomes detached from the $n=1,2$ rings, particularly in the receding (right) half of the image. Note that observations suggest that M87*'s SED peaks at $\nu>230$~GHz~\citep{EHTMWL}.

\begin{figure*}[]
        \centering
        \includegraphics[width=\textwidth]{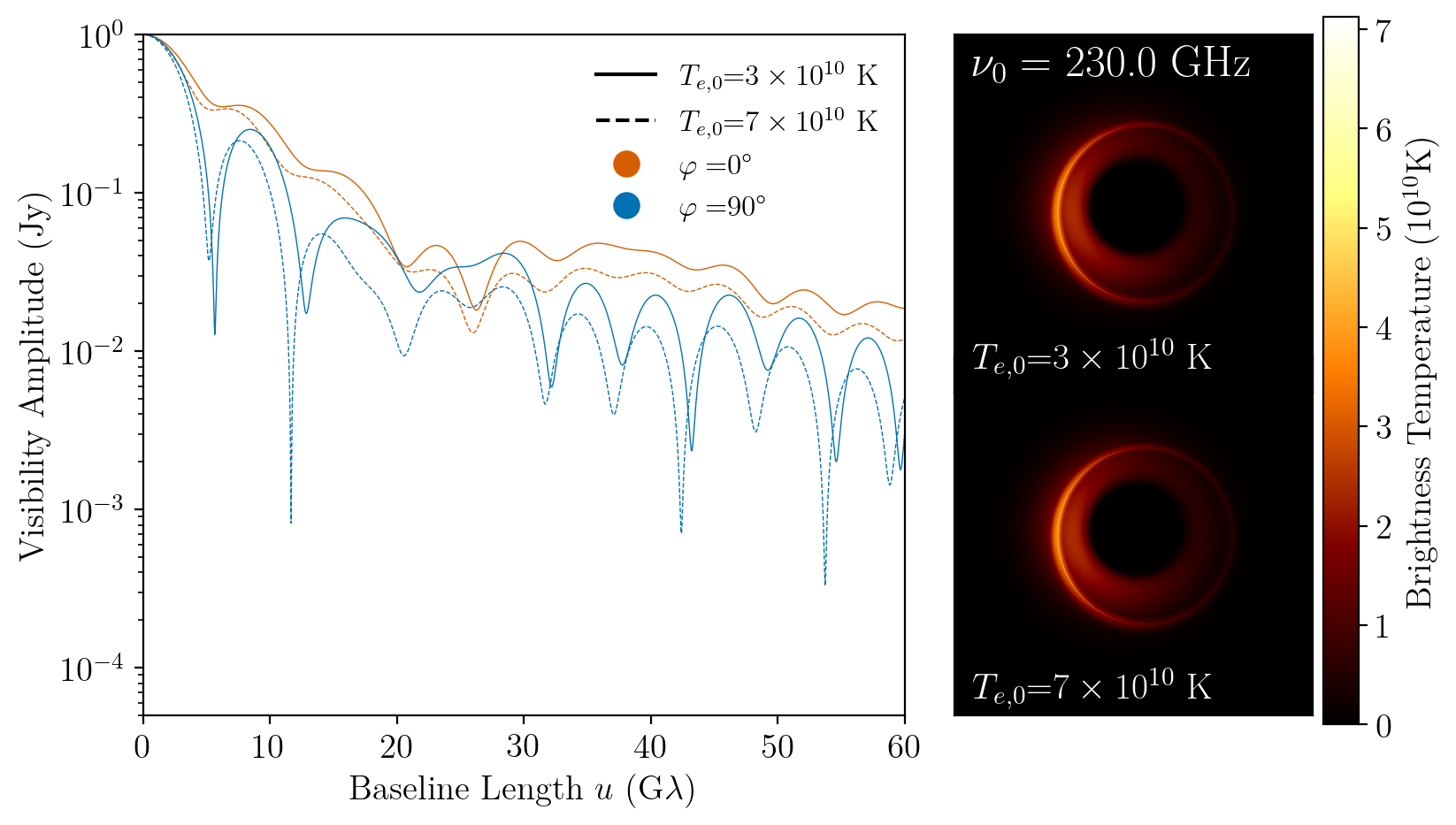}
        \vspace{5pt}
    \caption{Visibility amplitudes for two models at two baseline angles, $\varphi=0\degree$ (orange curves) and $\varphi=90\degree$ (blue curves), at $\nu_0=230$~GHz. Solid lines denote the case with $T_{e,0}=3 \times 10^{10}$~K, while the dashed lines indicate a case with $T_{e,0}=7 \times 10^{10}$~K. Variable parameters were fixed to $\alpha_T=1.0$, $\alpha_B=1.5$, $a_*=15/16$. All the remaining parameters are shown in \autoref{tab:constModels}. The corresponding images are shown on the right. As an alternative method to infer properties (as done in the image domain in  \autoref{fig:ThickRadVNuGrid}), one should also be able to infer an (interferometric) diameter from these ringing patterns \citep[e.g.][]{GrallaLupsascaMarrone2020}.}
        \label{fig:visamp}
    \end{figure*}

The combined contributions of $n=0,1,2$ images to the average radius defined in the image domain suggests that it may be better to analyze also these photon rings in the visibility domain, where different spatial scales are naturally separated~\citep{Cardenas-Avendano:2023dzo}. We leave such analyses for future work, and here provide examples of the visibility amplitudes for two models in \autoref{fig:visamp}.

    \begin{figure}[h]
        \plotone{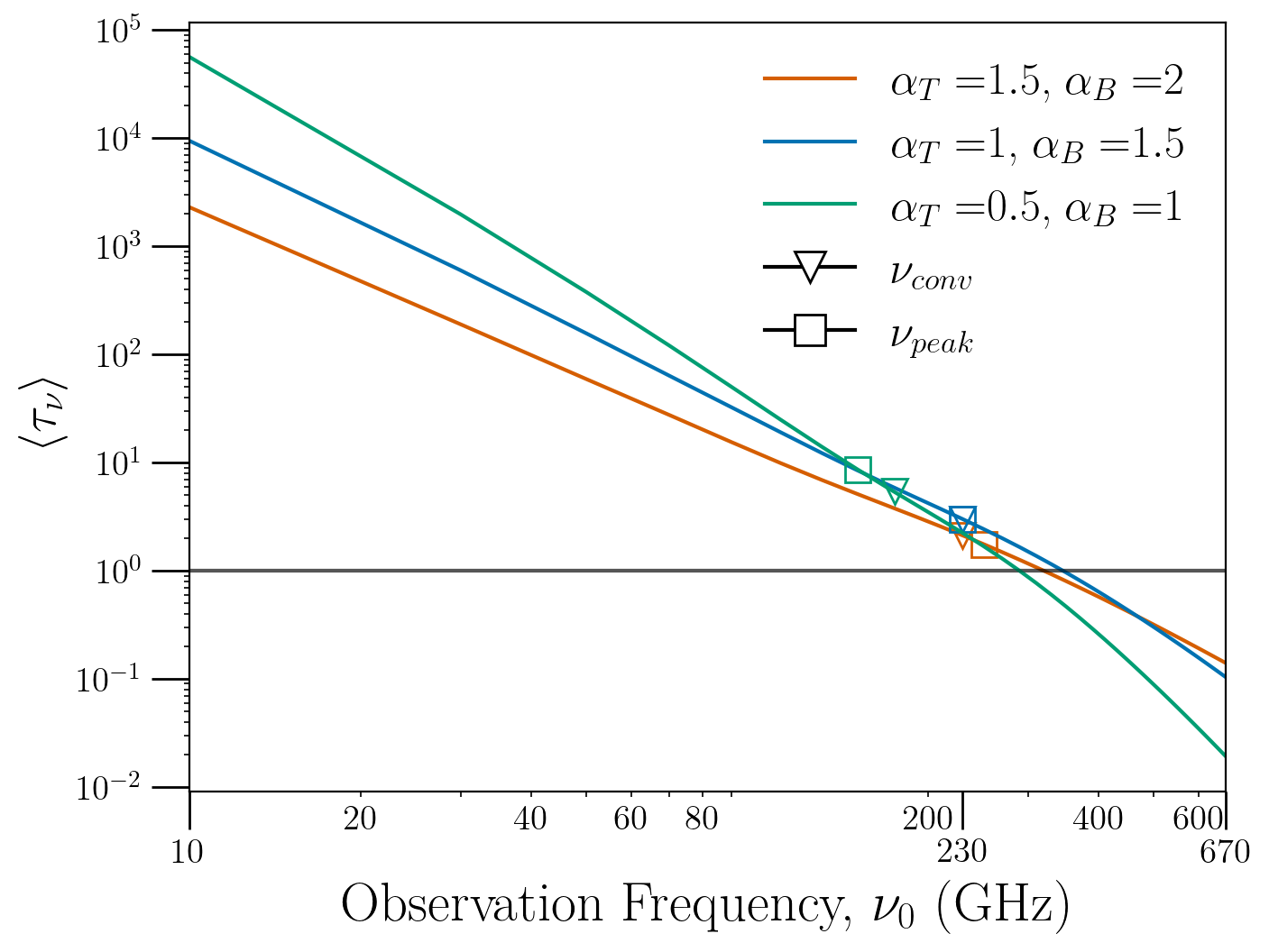}
        \centering
        \caption{The average optical depth (\autoref{eq:meantau}) as a function of the observation frequency. The triangles indicate $\nu_{\rm conv}$, a measure for the frequency at which the transition to the optically thin regime occurs defined in this work. The squares indicate $\nu_{\rm peak}$, i.e., the frequency at which the total flux density is maximized. All models were computed for a spinning black hole with $a_*=15/16$, $T_{e,0}=3 \times 10^{10}$~K, and the remaining parameter values as presented in~\autoref{tab:constModels}.
        }
        \label{fig:OpticalDepthGrid}
    \end{figure}
    
\subsection{Optical Depth \label{sec:OpticalDepth}}

To quantify the optical depth for each model, we compute
    \begin{align}
        \langle \tau_{\nu,n} \rangle = \frac{\int_{\rm Image} \tau_{\nu,n}  \mathcal{I}_{\nu_n}}{\int_{\rm Image}\mathcal{I}_{\nu_n}},
        \label{eq:meantau}
    \end{align}
i.e., a weighted average of the optical depth over the image for each ring. One can define the optically thin regime whenever $\langle \tau_{\nu,n} \rangle \leq 1$. The optical depths for some models are shown in \autoref{fig:OpticalDepthGrid}. For these examples, we can see that $\langle \tau_{\nu,n} \rangle = 1$ occurs after $\nu_{\rm peak}$ and $\nu_{\rm conv}$, suggesting that the entire image is not optically thin even at these transition frequencies. 
 
\section{Towards more realistic black hole images: blurring and time-dependence}
\label{sec:noisyblurr}

\begin{figure*}
        \centering
        \plotone{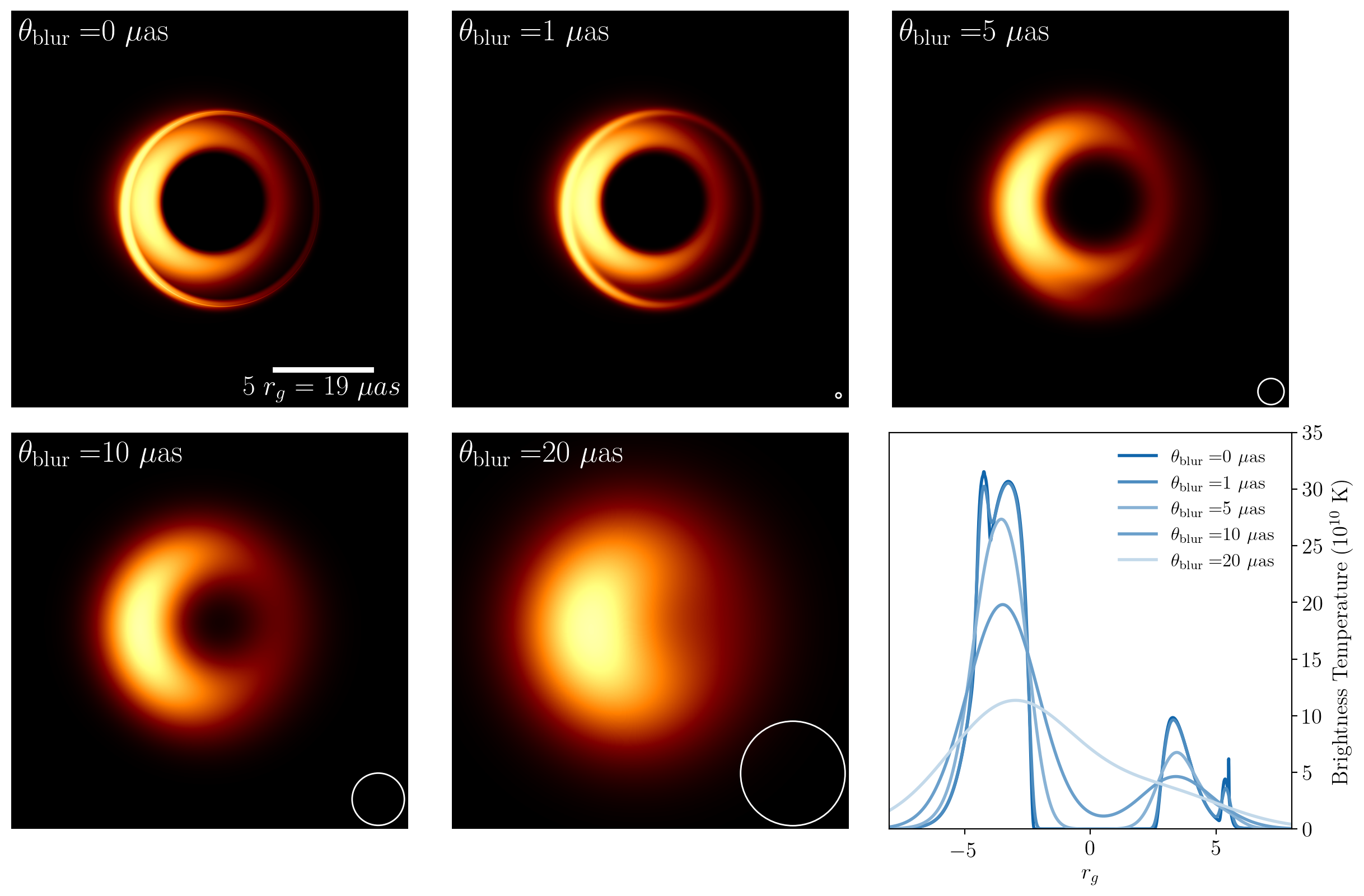}
        \caption{Blurring of M87* model images with different Gaussian smoothing kernels $\theta_{\rm blur}$, and their corresponding radial profiles (last panel). For all these cases the full RTE solution has been considered with $a_*=15/26$, $\alpha_T=1$, $\alpha_B=1.5$, $T_{e,0}=3 \times 10^{10}$~K, at $\nu_0=230$GHz. The remaining parameters are presented in \autoref{tab:constModels}. From the top left to the bottom right, the values of the Gaussian smoothing kernel FWHM are $\theta_{\rm blur}= [0,1,5,10]\,\mu$as. These are, respectively, approximate blurring kernels for ``infinite'' resolution, and 230 GHz baselines from Earth to L2 orbit,  from Earth to the Moon, from Earth to geostationary orbit (GEO), from Earth to low earth orbit (LEO), and EHT-like resolution. The inset circles indicate the FWHM diameter of the convolving kernel used for blurring the image. The last panel shows the radial emissivity curve for each image across the horizontal axes of the image.}
        \label{fig:blurring}
\end{figure*}

In the previous sections, we described images assuming ``perfect'' resolution. However, the analysis and characterization of these images are inherently tied to the resolution that can be realistically achieved in EHT observations, which is influenced by both the observation frequency and the baseline separations between telescopes. Gaussian blurring, in particular, plays a crucial role in this context, as it simulates the imperfections inherent in real-world observations. In ~\autoref{fig:blurring} we have applied different Gaussian blurring kernels, with full-width at half-maximum (FWHM) $\theta_{\rm blur}$,  to model images of M87* produced with our model. As expected, larger blurring kernels obscure the fine details of the higher $n$ rings, making the image more dominated by the already more spread (before blurring) $n=0$ contribution. A $\theta_{\rm blur} \leq 5$ corresponds to the expected resolution one can achieve with a low earth orbit (LEO) VLBI array telescope~\citep{JMD2020}. A similar baseline separation would also be needed to resolve the black hole inner shadow~\citep{Chael_Johnson_Lupsasca_2021}. 

When comparing models to real observations, it is essential to also consider the role of rapid time variability in the turbulent accretion flow close to the event horizon in M87. Following the prescription introduced in~\CA, we can easily incorporate a time dependence by generating a Gaussian random field and applying it to modify each power law. More specifically, we use realizations $GRF(x,y,t)$ of a Gaussian random field from the code \texttt{inoisy}~\citep{Lee_Gammie_2021}, 
to modify the disk power laws $f(r)$ in~\autoref{sec:diskModels} into functions that vary in space and time across the disk;
\begin{equation}
    f(x,y,t) = f(r) \times \exp\left[n_{\rm scale} \times GRF(x,y,t) - n_{ \rm scale} ^ {1 / 2}\right],
\end{equation}
where $n_{\rm scale}$ is a noise scale factor that controls the variability of the fluctuations. The time correlations of the underlying Gaussian field are prescribed to mimic physical properties, such as a Keplerian a flow with spiral-like structure as shown in~\CA. 

\begin{figure*}
            \centering
            \plotone{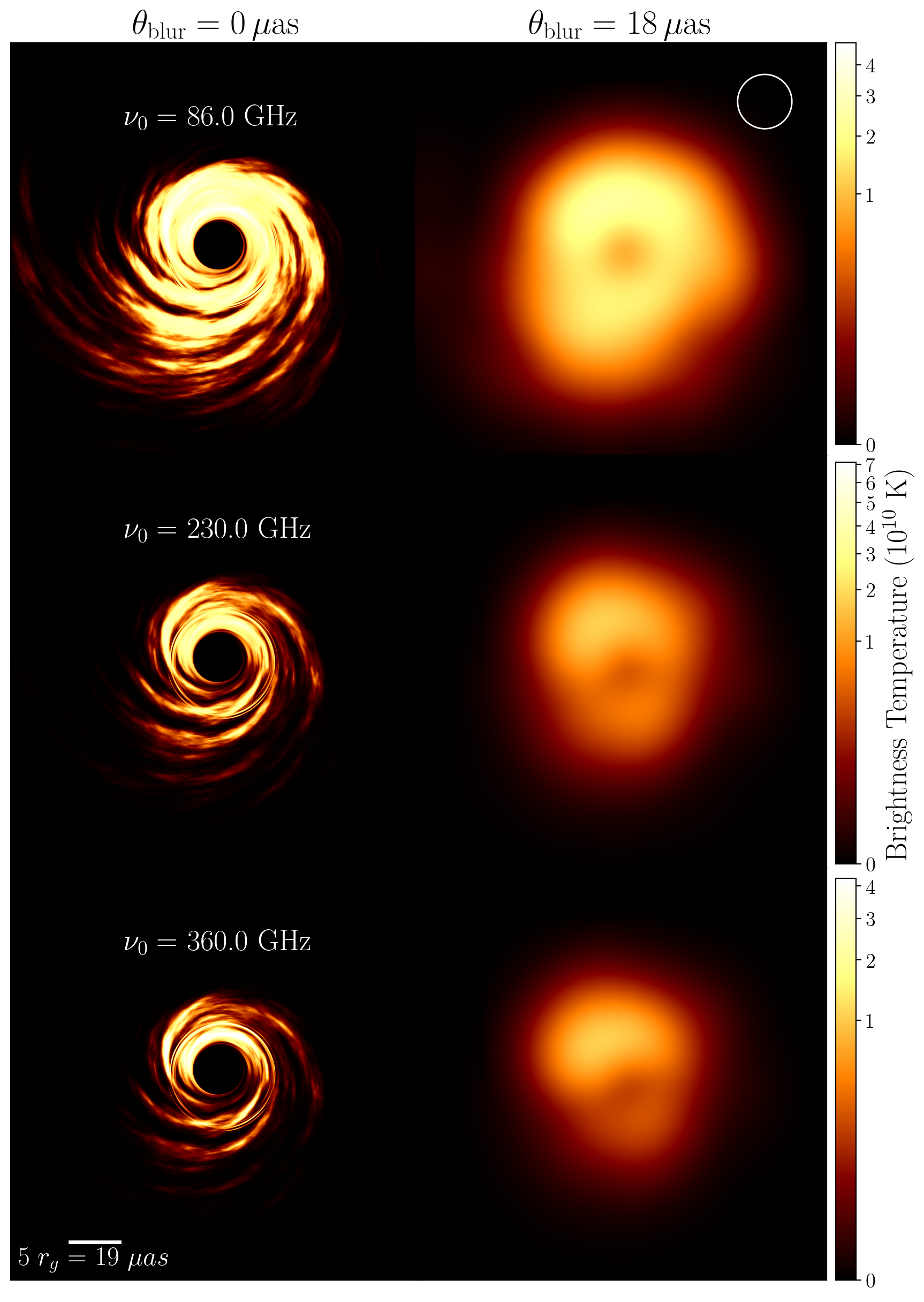}
            \caption{
            Snapshots at different frequencies  (left column) and their corresponding blurred version (right column). The variability and time dependence of the model has been added using \texttt{inoisy}. The right column images have been blurred using a Gaussian filter with a FWHM value of $18$ $\rm \mu as$. For all these images, $a_*=15/26$, $\alpha_T=1$, $\alpha_B=1.5$, $T_{e,0}=3 \times 10^{10}$~K, and noise scale $n_{\rm scale}=0.2$. The remaining parameters are shown in \autoref{tab:constModels}. Images on the same row share the same color bar. Note that the inclination angle in these images is $\theta_{\rm o}=17\degree$, and can be mapped to the $\theta_{\rm o}=163\degree$ for M87* by a flip in the image.} 
            \label{fig:Inoisy}
    \end{figure*}

Example snapshots of our model modified with this time variability implementation are shown in \autoref{fig:Inoisy}, where we also show the blurred version of each image (assuming $\theta_{\rm blur} = 10 \mu$as).
The difference in saturation levels shows that blurring the image reduces the overall brightness. This effect is expected to be independent of the observation frequency $\nu_0$. The previous extrapolations from the $\nu_0$ curves, which focused on their shapes rather than absolute heights, suggest that the previous analysis is still valid for black hole images that are not perfectly resolved, such as that for M87*.

A systematic study of the model parameters in the presence of both blurring and time variability to understand the parameter ranges for which the higher-order $n$ photon rings become visible is beyond the scope of the current work, and we plan to address it in future research. 

\section{Conclusions}
\label{sec:conclusion}

We have incorporated astrophysics-based emission profiles into the adaptive analytical ray-tracing code \texttt{AART}. The emission models are based on an analytic solution to the radiative transfer equation, enabling us to create high-resolution black hole images across a broad parameter space in an efficient way. For each model we analyzed the radii, flux density, and optical depth of the image and $n$ sub-images across observation frequencies. As expected, the size and shape of the higher $n$ sub-images are less dependent on astrophysical parameters. By injecting stochastic fluctuations into the disk power laws of our model and applying blurring with various-sized kernels, we have also shown how these models can be directly compared with images produced from GMRHD simulations and those captured by the EHT. 

In our ring radius measurements, we observe higher values of $n$ being influenced mainly by the black hole spacetime parameters. In contrast, the size and shape of the $n=0$ sub-image heavily depend on the assumed properties of the accretion disk. By comparing the average radius of the image's brightest point to the critical curve radius, we defined a frequency point, $\nu_{\rm conv}$ that, together with the peak flux observation frequency, $\nu_{\rm peak}$, serve as conservative estimates for the transition between the optically thick and thin regimes. We found that in most models with large black hole spin, the brightest point in the image on the receding/dim side of the image tracks $n=0$ even at high frequencies, so the average image radius lies between the $n=0$ and lensed $n=1$ and $n=2$ radii. At low spin, there is a range of image radii where the flow is both optically thin, and the $n=1$ and $n=2$ rings are the brightest points in the resolved image on both sides of the flow. 

Within the optically thick regime, the direct image ($n=0$) size is dictated mainly by the thermal electron temperature. In the optically thin regime, all astrophysical parameters of the model become crucial, with the the magnetic field strength being the dominant. This points out one advantage of multi-frequency analysis, as observations in both regimes would be necessary to disentangle the accretion disk's magnetic field and temperature details.
    
Our results complement the recent efforts in~\cite{Palumbo:2024jtz}, where polarized images of two models favored by the EHT were systematically studied. Since our approach is significantly less computationally expensive, it allowed us to explore a larger parameter space. Future work will be devoted to uncovering which points in the parameter space lead to more extreme outputs, making detailed comparisons to GRMHD simulations, and adding polarization capabilities and non-thermal electron and jet emissions. For instance, jet emission can be approximated as emission on a conical surface, as recently modeled in~\cite{Chang_Johnson_Tiede_Palumbo} with off-equatorial cones. 

Adding non-thermal electrons to our model may lead to a significantly more accurate flux density output at lower observation frequencies (sub 50 GHz) \citep{Broderick_Fish_Doeleman_Loeb_2011}. Although this may not be a relevant region to current EHT observations, accurate modeling of the lower frequency image will become important when the EHT includes simultaneous 86 GHz observations \citep{Chael_Issaoun_Pesce_Johnson_Ricarte_Fromm_Mizuno_2023} or when jointly modeling EHT data with data from lower-frequency VLBI. To further increase accuracy in the opposite end of the spectrum for observation frequencies above $1000$ GHz, we will need to add jet emission, which is shown to dominate at such ranges \citep{2009ApJ...697.1164B}. However, as shown in this work, the disk emission should have significantly decreased at those frequencies.

By blurring time-dependent images, we have demonstrated that our physical accretion disk model implemented in \texttt{AART} can be quickly extended to explore the effects of turbulent structure and time variability on the observables studied here. The methods introduced in this work aim to build a fast, flexible, but physically motivated model for black hole photon rings at multiple frequencies. Multi-frequency observations of black hole accretion flows can disentangle the complex interplay between astrophysical processes and spacetime geometry, providing explicit constraints on black hole mass and spin. Simulating these systems will help shape the development of future observation campaigns and missions such as the Black Hole Explorer satellite \citep{BHEX}, turning these astrophysical sources into precise laboratories for both gravitational and astrophysical phenomena.

\begin{acknowledgments}

We thank Eliot Quataert and Daniel Palumbo for their useful comments. This work used resources provided by Princeton Research Computing,
a consortium that includes PICSciE (the Princeton Institute for Computational Science and Engineering) as well as the Office of Information Technology’s Research Computing division. A.C.-A. acknowledges support from the US Department of Energy through the Los Alamos National Laboratory. Additional funding was provided by the Laboratory Directed Research and Development Program, and the Center for Nonlinear Studies at Los Alamos National Laboratory under project number 20240748PRD1. This article is cleared for unlimited release (LA-UR-24-31211). AC was supported by the Princeton Gravity Initiative. 
\end{acknowledgments}

\software{astropy \citep{2022ApJ...935..167A, 2018AJ....156..123A, 2013A&A...558A..33A}.  \text{AART} \CPA. \texttt{Inoisy} \citep{Lee_Gammie_2021}. \texttt{Numpy} \citep{harris2020array}. \texttt{Lmfit} \citep{2016ascl.soft06014N}}

\appendix

\section{Model Normalization}
\label{appendix:norm}
 
We normalize all models of M87* to produce a total flux density of $0.5$ Jy at 230 GHz. We chose $n_{th,0}$ as the normalization variable. To determine $n_{th,0}$, we create a function $G(n_{th,0})=\left[0.5 - F_\nu(n_{th,0})/ {\rm  1 Jy}\right]$ for images produced at $230$~GHz. We evaluate $G(n_{th,0})$ at different values of the density normalization using a lower resolution ($250 \times 250$ pixels, corresponding to an image computed on a range of $[-25\,r_{\rm g},25\,r_{\rm g}]$ with a resolution of $0.2\,r_{\rm g}$). Lastly, using \texttt{lmfit}'s method of least squares in the \texttt{minimize} function, we find the value of $n_{th,0}$ that minimizes $G(n_{th,0})$~\citep{2016ascl.soft06014N}. The reference total flux density of $\sim0.5$ $\rm Jy$ is motivated by M87* observations at $230$~GHz \citep{EHTM87IV}. The $n_{th,0}$ values for all $81$ models ranged from $2.8 \times 10^{3}$ $-$ $4 \times 10^{6}$ $\rm cm^{-3}$, having a mean of $4 \times 10^{5}$ $\rm cm^{-3}$ and a standard deviation of $8.8 \times 10^{5}$ $\rm cm^{-3}$.

\section{Convergence of the images: A resolution study}
\label{appendix:convergence}

\begin{figure}[]
    \centering
    \includegraphics[width=200pt]{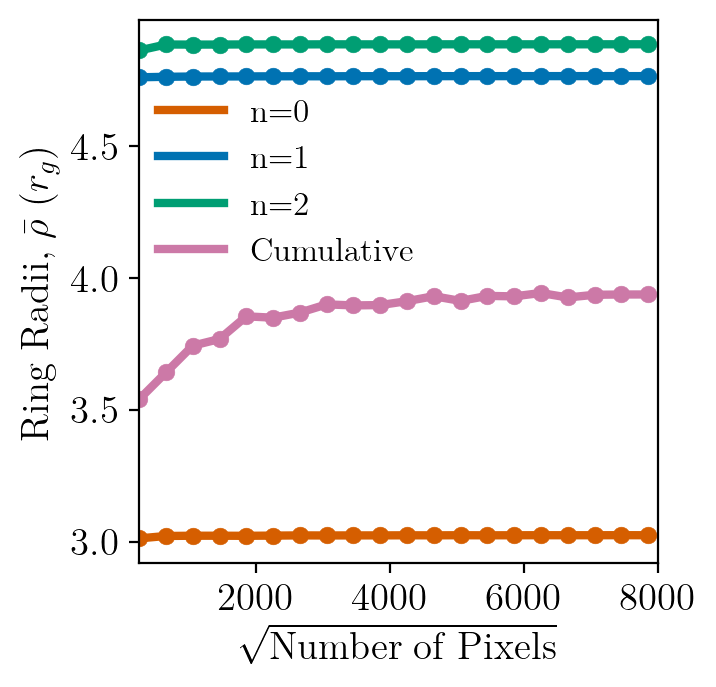}
    \caption{An example of the resolution convergence test performed for the inferred radii in this work. For this example, we have chosen a model with $a_*=15/26$, $\alpha_T=1$, $\alpha_B=1.5$, $T_{e,0}=3 \times 10^{10}$~K, $\nu_0=670$ GHz, and the remaining model parameters as shown in \autoref{tab:constModels}. The higher the resolution, the more accurate the inference of these radii. The individual photon ring radii become converged roughly at a resolution of $1000\times1000$ pixels, while the radius of the cumulative image converges more slowly. This asymptotic approach prevents a cleanly defined point in which all calculations exactly become independent of resolution. At a resolution of $8000\times8000$ pixels the differences in the inferred radius is below $1\times10^{-4}\,{r_{\rm g}}$ with respect to the image at $7500\times7500$ pixels.}
    \label{fig:convTest}
\end{figure}

To ensure that the image radii calculated in \autoref{sec:ringSizes} are independent of image resolution, we computed all the images at various resolutions (linearly distributed according to pixel size) and measured their radii as a function of the number of image pixels. After plotting the behavior of the ring radii with resolution in~\autoref{fig:convTest}, we chose the value $8000\times8000$ pixels as our fiducial resolution. In~\autoref{fig:convTest} we see that our chosen resolution is at a point where the cumulative radii has not become completely independent of resolution, but the changes in the inferred radius are less than $1\times10^{-4}\,{r_{\rm g}}$. 

\begin{figure}[ht!]
    \plotone{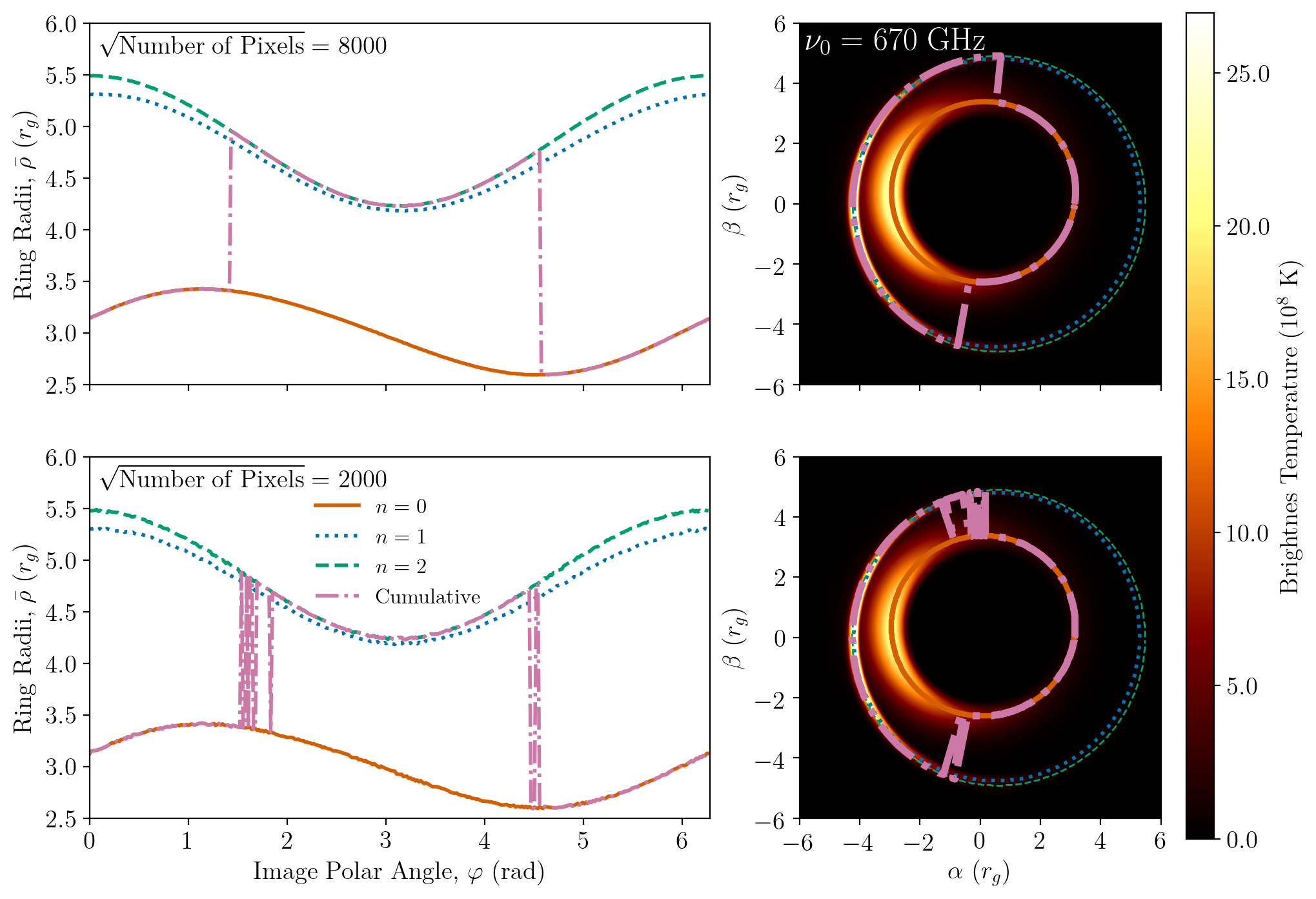}
    \caption{The dependence of the measured photon ring radii, as a function of the polar angle $\varphi$ for two different resolutions. For this example images, the model parameters were set to $a_*=15/16$, $\alpha_T=1$, $\alpha_B=1.5$, $T_{e,0}=3 \times 10^{10}$~K, with the remaining parameters as listed in~\autoref{tab:constModels}. The top row is produced on a grid resolution of $\sqrt{\rm Number \; of \; Pixels}=8000$ pixels, our nominal resolution, well within the range of being safely independent of pixel resolution, as seen in~\autoref{fig:convTest}. The bottom row is produced on a lower resolution with $2000\times2000$ pixels. The top row's single jump of the cumulative radii from $n=0$ to $n=2$ and back to $n=0$ is physically motivated, as discussed in \autoref{sec:ringSizes}. On the other hand, the bottom row's multiple jumping (switching) is not physical, and results because the correct peaks for $n=2$ are not being captured. This effect is particularly important for the $\varphi$ regions in which ring $n=2$'s intensity is close to $n=0$'s.}
    \label{fig:rollingWindow}
\end{figure}

As we can see from \autoref{fig:convTest}, the radius of the cumulative image requires a higher resolution to converge than the radius of any individual sub-image. This is likely mainly due to the failure of lower resolution to capture the actual intensity peak for the narrow $n=2$ sub-image within some pixels. When the width of the (very thin) $n=2$ ring is smaller than the size of a single pixel, the actual peak brightness of the $n=2$ ring may not be captured. This does not affect the radius of the $n=2$ image alone, but it becomes significant around the $\varphi$ where the highest intensity of the cumulative image transitions from $n=0$ to $n=2$ (as seen in the ``jumping'' between sub-images in \autoref{fig:670GHZ}). Therein, the difference in intensities for the $n=2$ and $n=0$ sub-images is very small, so a loss of some intensity in $n=2$ may cause the $n=0$ peak to be more prominent, switching the cumulative radius to $n=0$. This effect depends only on how the ring happens to be placed under the pixel grid, meaning that the peak intensity of the cumulative image radial slice will switch back and forth between $n=2$ and $n=0$. To avoid a non-physical switching, higher resolutions are needed to accurately capture the $n=2$ peak intensities per pixel. As an example, in \autoref{fig:rollingWindow} we show the same model and inferred radii for two different resolutions. As we can see from the top panel therein, the ``switching'' artifact does not occur for sufficiently high-resolution images. For all the models studied in this work, the nominal resolution ($8000\times8000$) was enough to prevent these artifacts from affecting our inferred ring radii measurements. 

\FloatBarrier
\bibliographystyle{aasjournal}
\bibliography{main}

\end{document}